\documentclass[pra,twocolumn,showpacs,superscriptaddress]{revtex4-1}
\def\iota{\imath}
\usepackage[psamsfonts]{amssymb}
\usepackage{amsmath}
\usepackage{epsf,epsfig}
\usepackage{graphicx}
\usepackage{lipsum}
\begin{document}
\title{Spontaneous emission and the operation of invisibility cloaks: Can the invisibility cloaks render objects invisible in quantum mechanic domain?
\\}
\author{Mina Morshed Behbahani}
\affiliation{Department of Physics, Faculty of Basic Sciences,
Shahrekord University, P.O. Box 115, Shahrekord 88186-34141, Iran.}
\author{Ehsan Amooghorban}
\email{Ehsan.amooghorban@sci.sku.ac.ir} \affiliation{Department of
Physics, Faculty of Basic Sciences, Shahrekord University, P.O. Box
115, Shahrekord 88186-34141, Iran.} \affiliation{Photonics Research
Group, Shahrekord University, P.O. Box 115, Shahrekord 88186-34141,
Iran.}
\author{Ali Mahdifar}
\affiliation{Department of Physics, Faculty of Basic Sciences,
Shahrekord University, P.O. Box 115, Shahrekord 88186-34141, Iran.}
\affiliation{Photonics Research Group, Shahrekord University, P.O.
Box 115, Shahrekord 88186-34141, Iran.}

\begin{abstract}
As a probe to explore the ability of invisibility cloaks to conceal objects in the quantum mechanics domain, we study the spontaneous emission rate of
an excited two-level atom in the vicinity of an ideal invisibility cloaking. On this base, first, a canonical quantization scheme is presented for the
electromagnetic field interacting with atomic systems in an anisotropic, inhomogeneous and absorbing magnetodielectric medium which can suitably be
used for studying the influence of arbitrary invisibility cloak on the atomic radiative properties.
The time dependence of the atomic subsystem is obtained in the Schrodinger picture. By introducing a modified
set of the spherical wave vector functions, the Green tensor of the system is calculated via the continuous and discrete methods. In this formalism, the
decay rate and as well the emission pattern of the aforementioned atom are computed analytically for both
weak and strong coupling interaction, and then numerically calculations are done to demonstrate the performances of cloaking in the quantum mechanics
domain. Special attention is paid to different possible orientations and locations of atomic system near the spherical invisibility cloaking. Results
in the presence and the absence of the invisibility cloak are compared.
We find that the cloak works very well far from its resonance frequency to conceal a macroscopic object, whereas at near the resonance frequency the
object is more visible than the situation that the object is not covered by the cloak.
\end{abstract}

\maketitle

\section{Introduction}
As a result of the implement of many intriguing features that have not yet been found
in nature, metamaterial has attracted a great deal of attention in the field of optics. These features prepare new opportunities for realizing exotic phenomena such as invisibility devices~\cite{Pendry 2006}-~\cite{Schurig 2006}, superlenses~\cite{Pendry 2000,Schurig 2007}, field rotators~\cite{Chen 2007}, optical analogues of black holes~\cite{Narimanova 2009, Genov 2009}, Schwarzschild spacetime~\cite{Chen 2010}, wormholes~\cite{Greenleaf 2007} and the "Big Bang" and cosmological in
ation~\cite{Smolyaninov 2011,Smolyaninov 2012}. The theoretical basis for some of these phenomena is coordinate transformation, which stems
from the formal invariance of Maxwell's equations. This enable both physics and engineering societies to manipulate electromagnetic waves in almost
any
fashion. In this paper, we focus on the invisibility cloaking, and attempt to gain some physical insight regarding the quantum electrodynamics of this
topic.

Based on a coordinate transformation, Pendry et al. first proposed an invisibility cloak,
which can protect the cloaked object of arbitrary shape from
electromagnetic radiation. The external observer got therefore unaware of the presence of the cloak and the object. This idea has been verified
numerically by full-wave simulations~\cite{Cummer 2006} and experimentally by using metamaterial at the microwave frequency~\cite{Schurig 2006}, and even at the optical
frequencies~\cite{Cai 2007}. However, these invisibility cloaks were encountered a serious limitation: they required extreme values of material properties and
can
only work within a narrow-band frequency. The first issue is circumvented by using simplified constitutive parameters~\cite{Schurig 2006}. To overcome the bandwidth limitation, it was proposed carpet cloak to conceal an object that is placed
under a bulging reflecting surface by imitating the reflection of a flat surface~\cite{Li 2008}. Such cloaks were experimentally demonstrated in both microwave~\cite{Liu 2009,Ma 2010} and optical frequencies~\cite{Valentine 2009}-\cite{Ergin 2010} using metamaterial structures with feature sizes in the centimeter to nanometer scale. It inevitably requires complicated nanofabrication processes which restrict the size of hiding objects in the visible frequencies to a few wavelengths. So, this carpet cloaking has not enabled to hide a large object at least as experimentally. Furthermore, the scattered waves are suffered a lateral shift, which makes the object to be detectable. To bypass these limitations, macroscopic invisibility cloaking has been introduced by using the birefringence property of a natural crystal such as calcite at broadband visible
wavelengths~\cite{Chen 2010,Zhang 2011}.
%

As mentioned above briefly, all researches on cloaking reported
so far have been focused on the case in which the electromagnetic field is taken into account classically. However, the performances of cloaking in the quantum mechanics domain, which can be an important topic, have been rarely considered~\cite{Kamp 2013}. The question that naturally arises in this context is
whether such invisibility clocks work in the quantum mechanics domain as well as in the classical regime. The present paper is intended to respond to
this question.
Without loss the generality of our approach, we restrict our attention to the special case at Pendry cloaking which
was relatively easy to be constructed, simulated and analyzed.

It is well known that the spontaneous decay of an
atom is influenced by the geometry and the optical properties of the material
body. We therefore expect that if an atom is located near the ideal invisibility cloaking, the electromagnetic interaction between them will lead to
drastically modification of the density of radiation modes
and subsequently the spontaneously decay rate. In this sense, it is
useful to look at the decay rate of an excited atom as a
probe that allows us to examine the operation of this type of cloaks in the quantum mechanic regime.
On this base, we study the spontaneous emission of an exited two-level atom and as well the spatial distribution of its radiation intensity as a
function of the atomic transition frequency and the distance between the atom and the invisibility cloaking.

On one hand, the spontaneous emission is a phenomena which corresponds to the conventional framework of quantum electrodynamics. On the other hand,
such invisibility cloak with position-dependent and anisotropic optical parameters was demonstrated for microwave frequencies by utilizing concentric
layers of split-ring resonators.
%
%
Due to the metallic nature of resonator structure,
%
%
such invisibility cloaks are usually associated with a high loss factor and accompanied by a strong dispersion to fulfill
causality. Therefore, a fully quantum mechanical treatment is needed to
consider the dissipative and dispersive effects of the invisibility cloak along with their inhomogeneous and anisotropic features on the spontaneous
emission rate.

There are two approaches to quantize the electromagnetic field in the presence of a dissipative and dispersive media, in general: canonical and
phenomenological method. In this paper, we consider the rigorous canonical approach.
For this purpose, as a lateral purpose of the present paper, we extend the canonical quantization scheme in~\cite{Kheirandish 2010}-\cite{Amooghorban 2016} to a more general case that the
medium is described in terms of a spatially varying and anisotropic
permittivity and permeability, and as well the composed field-medium system interacting with atomic systems. This derivation helps us to gain a
physical insight into the influence of arbitrary cloak on the atomic radiative properties.

The paper is organized as follows. In Sec.~II we present a canonical quantization of the electromagnetic field interacting with charged particles in
presence of an anisotropic, inhomogeneous and absorbing magnetodielectric medium.
In Sec.~III, the spontaneous emission rate of an exited atom near a spherical invisibility cloak is expressed in terms of the imaginary part of the
classical Green tensor at the position of the atom. By expanding the Green tensor of system into a modified
set of the spherical wave vector functions, the decay rate is computed analytically in both
weak and strong coupling regime.
As an application of this formalism, the numerical evaluations are performed for the spherical cloak whose material absorption and dispersion is of the Lorentz type. Then, we discuss the role of the orientation of the dipole moment
of atomic system.
In Sec.~IV the spatially intensity of the spontaneously emitted
light is calculated as functions of the atomic transition frequency and the distance between
the atom and the hidden object which is covered by the spherical invisibility cloaking.
A summary and conclusions are presented in Sec. V. The derivation of the Green tensor via two methods, exact and discrete, is provided in
Appendix~\ref{Appen:Green tensor}.

\section{Canonical quantization of electromagnetic field }
Our analysis of the spontaneous decay of an excited atom placed in
vicinity of a Pendry cloaking is based on a generalization of a
canonical scheme for quantization of the electromagnetic field in an
isotropic magnetodielectrics medium developed in Refs~\cite{Kheirandish 2010}-\cite{Amooghorban 2016}. We
give only the bare essentials needed for an appreciation of the
present paper.

Based on an inspiration of the microscopic
Hopfield model~\cite{Hopfield 1958}, quantum electrodynamic in an inhomogeneous, anisotropic, dissipative
and dispersive magnetodielectric medium can be accomplished by
modeling the medium by two independent reservoirs comprised of a
continuum of three dimensional harmonic oscillators. These two independent sets of harmonic oscillators
are characterized by means of two harmonic oscillator fields
$\bf{X}_{\omega} $ and $ \bf{Y}_{\omega}$ which interact with the
electric and the magnetic fields through a dipole interaction term. Hereby, we can describe the polarizability and the magnetizability characters of
the
magnetodielectric medium, as well as its dissipative behavior.
Bearing these in mind, let us start with the total Lagrangian
density of the system composed of the electromagnetic
field, the external charged particles and the medium including the dissipative
behavior
\begin{equation}\label{total Lagrangian}
{\cal L} = {\cal L}_{\rm EM} +  {\cal L}_{\rm e} +  {\cal L}_{\rm m}
+  {\cal L}_{\rm q} + {\cal L}_{\rm int},
\end{equation}
where the electromagnetic part ${\cal L}_{\rm EM}$ has the standard
form ${\cal L}_{EM} = \frac{1}{2} \varepsilon_{0} {\bf E}^{2}({\bf
r} , t )  - \frac{1}{2\mu_{0}} {{\bf B}^{2}}({\bf r },t)  $ that the
electric field ${ \bf E}= - {\partial {\bf A}} / {\partial t} -  {
\bf \nabla} \varphi $ and magnetic field $ \bf{B} = \bf{\nabla}
\times \bf{A} $ are written in terms of the vector potential $ \bf A
$ and scalar potential $ \phi $.
The electric and the magnetic parts of the material Lagrangian
density ${\cal L}_{e}$ and ${\cal L}_{m}$, which are modeled by a
continuum of harmonic oscillators, are given by $\frac{1}{2} \int_
{0}^{\infty} \mbox{d}\omega \left[ \dot{{ \bf X}}^{2}_{\omega}({\bf
r} ) - \omega^{2}{ \bf X}^{2}_{\omega}({\bf r}  ) \right]$ and $
\frac{1}{2} \int_ {0}^{\infty} \mbox{d}\omega \left[ \dot{{\bf
Y}}^{2}_{\omega}({\bf r} ) - \omega^{2}{\bf Y}^{2}_{\omega}({\bf r}
) \right]$, respectively. The polarization and magnetization fields
of the medium in term of two harmonic oscillator fields
$\bf{X}_{\omega} $ and $ \bf{Y}_{\omega}$ are defined as
\begin{subequations}\label{P and M field}
\begin{eqnarray}
{\bf P}({\bf r} , \omega ) = \int _{0}^{\infty}\mbox{d}\omega\,\,\, \bar{ \bar g}_{e} ({\bf r} , \omega ) \cdot {\bf X}_{\omega}({\bf r} , \omega
),\label{polarization field}\\
{\bf M}({\bf r} ,\omega ) = \int _{0}^{\infty}\mbox{d}\omega\,\,\, \bar{
\bar g}_{m} ({\bf r} , \omega ) \cdot{\bf Y}_{\omega}({\bf r} ,
\omega ), \label{magnetization field}
\end{eqnarray}
\end{subequations}
where the interaction with the material is described via the
coupling tensors, $ \bar{ \bar g}_{e} ({\bf r} , \omega ) $ and $
\bar{ \bar g}_{m} ({\bf r} , \omega ) $, which are assumed to be
analytic functions of $\omega$ in the upper half
plane. It is worth noting that the permittivity and the permeability of medium will be determined in term of these coupling tensors. So, we take here
the coupling tensor as a function of position to be the second order tensor, since the Pendry clocks under consideration are nothing but an
inhomogeneous and anisotropic metamaterial.\\
The forth term in Eq.~(\ref{total Lagrangian}) is the Lagrangian
density of free charged particles with particles mass $ m_{\alpha} $
and position $ {\bf r} _{\alpha} $, which is written as
\begin{equation}\label{charged particle Lagrangian}
{\cal L}_{q} =\frac{1}{2}\sum_{\alpha} m_{\alpha}\dot{{\bf
r}}_{\alpha}^{2}.
\end{equation}
Finally, ${\cal L}_{int} $ is the interaction Lagrangian density
which includes the linear interaction between the medium and the
charged particles with the electromagnetic field. It is found that
such interaction is given by
\begin{eqnarray}\label{interaction Lagrangian}
{\cal L}_{int}&=&{\bf J}({\bf r}_{\alpha} , t )\cdot{\bf A}({\bf r}_{\alpha} , t ) - \rho({\bf r}_{\alpha}) \varphi({\bf r}_{\alpha})\nonumber \\
&+&{\bf P}({\bf r} , t )\cdot {\bf E}({\bf r}, t ) + {\bf{ \nabla}}
\times {\bf A}({\bf r} , t )\cdot {\bf M}({\bf r} , t)
\end{eqnarray}
where $ {\bf J}({\bf r}_{\alpha} , t ) $ is the current density of
charged particles. An analysis of the Lagrangian density~(\ref{total
Lagrangian}) shows that the scalar potential $\dot{\phi}$ does not
appear. Therefore, the scalar potential is not a proper dynamical
variable and the corresponding equation of motion can be treated as
a constraint. It enables us to eliminate the scalar potential, and
then get a reduced Lagrangian where only the vector potential {\bf A},
the material fields ${\bf X}_\omega$ and ${\bf Y}_\omega$, and their
time derivatives are appeared. To do this, we apply Euler-Lagrange equations
to the scalar potential. It leads to
\begin{eqnarray}\label{potential A+P}
{\varphi} &=& \varphi_{A}+\varphi_{p}\nonumber\\
&=&\frac{1}{4\pi\varepsilon_{0}}\int  \mbox{d}^{3} r^{\prime}
\frac{{\rho}_{A} (\bf r^{\prime})} {\vert {\bf r } - {\bf
r^{\prime}} \vert } + \frac{1}{4\pi\varepsilon_{0}} \int  \mbox{d}^{3}
r^{\prime} \frac{{\rho}_{p}(\bf r ^{\prime})} {\vert {\bf r } - {\bf
r^{\prime}} \vert },
\end{eqnarray}
where ${\rho}_{A} (\bf r) = \sum_ {\alpha} e_{\alpha} \delta( r -
r_{\alpha})$ and ${\rho}_{p}(\bf r) = -\nabla \cdot {\bf P}(\bf r)$
are the charge density and polarization-charge density,
respectively, and subsequently $ \varphi_{A} $ and $ \varphi_{P} $
are the corresponding scalar potentials arisen from these charge
distributions. By substituting Eq.(\ref{potential A+P}) into
Lagrangian~(\ref{total Lagrangian}), the total Lagrangian can be
recast into the reduced form
\begin{eqnarray}\label{final total Lagrangian}
{\cal L}&=& \frac{1}{2}\sum_{\alpha} m_{\alpha}\dot{{\bf r}}_{\alpha}^{2} + \frac{1}{2}\,\varepsilon_{0} {\dot{ \bf A}}^{2}({\bf{r}},t) -
\frac{1}{2\mu_{0}}\left(\nabla \times {\bf A}({\bf{r}},t)\right)^{2} \nonumber \\
&+&\frac{1}{2} \int_ {0}^{\infty} \mbox{d}\omega \left\lbrace {\dot {\bf X }}^{2}_{\omega}({\bf{r}},t) - \omega ^{2} {\bf X
}^{2}_{\omega}({\bf{r}},t)\right\rbrace \nonumber \\
&+& \frac{1}{2} \int_ {0}^{\infty} \mbox{d}\omega \left\lbrace {\dot {\bf Y }}^{2}_{\omega}({\bf{r}},t) - \omega ^{2} {\bf Y
}^{2}_{\omega}({\bf{r}},t)\right\rbrace  \nonumber \\
&+& \sum_{\alpha} e_{\alpha} \dot {\bf r}_{\alpha} \cdot {\bf A}
({\bf r}_{\alpha},t) + {\bf A} \cdot \dot {\bf P}({\bf{r}},t)+{\bf
M}\cdot \nabla \times  {\bf
A}({\bf{r}},t)\nonumber \\
&-& W_{coul},
\end{eqnarray}
where $ W_{coul} $ is Coulomb energy of the charged particles, the
polarization-charge and their interactions which in term of $
\varphi_{A} $ and $ \varphi_{P} $ is defined as
\begin{eqnarray}\label{W _{coul}}
W _{coul} &=& \frac{1}{2} \int  \mbox{d}^{3} r {\rho}_{A} ({\bf
r})\, {\varphi }_{A}({\bf r}) +  \int  \mbox{d}^{3} r {\rho}_{A}
({\bf r})\, {\varphi
}_{P}({\bf r})\nonumber \\
 &+&\frac{1}{2} \int  \mbox{d}^{3} \,r {\rho}_{P} ({\bf
r})\,{\varphi }_{P}({\bf r}).
\end{eqnarray}
The Lagrangian~(\ref{final total Lagrangian}) can now be used to
obtain the corresponding canonical conjugate variables for the
fields
\begin{subequations}\label{canonicall conjugate variables}
\begin{eqnarray}
&&\hspace{-0.5cm}- {\varepsilon _0}{\bf{E}}^{\bot}({\bf{r}},t) = \frac{{\delta L}}{{\delta \dot{\bf{A}}({\bf{r}},t)}} = {\varepsilon _0} \dot{{\bf A}}({\bf{r}},t),
\label{canonicall conjugate variables E} \\
&&\hspace{-0.5cm}{\bf Q }_\omega ({\bf{r}},t) = \frac {{\delta L}} { \delta {\dot{{\bf X}}_{\omega }({\bf r},t)}}= {\bar{\bar{g}}_{e}}({\bf{r}},\omega
){\bf{A}}({\bf{r}},t) +\dot{{\bf X }}_{\omega } ({\bf{r}},t),\label{canonicall conjugate variables Q} \\
&&\hspace{-0.5cm}{{\bf \Pi} _\omega }({\bf{r}},t) = \frac{{\delta L}} { \delta {\dot{{\bf Y }}_{\omega }({\bf r},t)}} =  {\dot{\bf{Y}}}_\omega
({\bf{r}},t),\label{canonicall conjugate variables Pi} \\
&&\hspace{-0.5cm}{\bf p}_{\alpha}({\bf r}_{\alpha} , t ) = \frac{\partial
L}{\partial {\dot{r_{\alpha}}}} = m_{\alpha}{\bf r}_{\alpha} +
e_{\alpha} {\bf A}({\bf r}_{\alpha} , t )\label{canonicall conjugate
variables p}.
\end{eqnarray}
\end{subequations}
Now, the transition from the classic to the quantum domain can be
accomplished in a standard fashion by applying commutation relation
on the variables and their corresponding conjugates. For
the electromagnetic field, we have
\begin{eqnarray}\label{commutation relation for A}
\left[ {\hat{A}} ({r} , t ), - \varepsilon_{0}{\hat{ E}}^{\perp} ({
r^{\prime}} , t )\right] &=& i \hbar \delta^{\perp}({ r} - {
r^{\prime}}),
\end{eqnarray}
and for the material fields and the dynamical variable of charged
particles
\begin{subequations}\label{commutation relation for XYp}
\begin{eqnarray}
\left[ {\hat{X}}_{\omega} ({r} , t ), {\hat{Q}}_{\omega^{\prime}} ({ r^{\prime}} , t )\right] &=&  i \hbar \delta ({r} - { r^{\prime}}) \delta
({\omega
} - \omega^{\prime}),\label{commutation relation for X} \\
\left[ {\hat{Y}}_{\omega} ({ r} , t ), \Pi_{\omega^{\prime}} ({ r^{\prime}} , t )\right] &=&  i \hbar \delta ({r} - {r^{\prime}}) \delta ({\omega } -
\omega^{\prime}),\label{commutation relation for Y} \\
\left[ q_{\alpha},{\hat{ p}}_{\beta}({r} , t )\right] &=& i \hbar
\delta_{\alpha\beta} \label{commutation relation for p}.
\end{eqnarray}
\end{subequations}
By applying the Lagrangian~(\ref{total Lagrangian}) and the expressions
for canonical conjugate variables in~(\ref{canonicall conjugate
variables}), we can form the Hamiltonian density as,
\begin{eqnarray}\label{Hamiltonian}
{\cal H} &=& \sum _\alpha  \frac{1}{2{m_\alpha }}\left[ {{\bf{p}}_\alpha }({\bf r}_\alpha ,t) - e_{\alpha }A ({\bf r}_{\alpha },t) \right] ^2 +
\frac{1}{2}{\varepsilon _0}{\bf{E}}^{\bot 2} ({\bf{r}},t)\nonumber \\
&+& \frac{{{\bf{B}}^2}({\bf{r}},t)} {2 \mu _{0}} + \frac{1}{2} \int_0^\infty  \mbox{d} \omega \, \left\{{\bf{Q}} _\omega ({\bf{r}},t) + {\omega ^2}\dot
{\bf{X}}_\omega ^2({\bf{r}},t) \right\} \nonumber \\
&+&\frac{1}{2}\int_0^\infty  \mbox{d} \omega \left\{{ {\bf \Pi}} _\omega ({\bf{r}},t) + {\omega ^2} \dot {\bf Y} _\omega ^2({\bf{r}},t) \right\} \nonumber \\
&-& {\bf \nabla} \times {\bf A}({\bf{r}},t)\cdot \,{\bf{M}}({\bf{r}},t) -\dot{\bf P }({\bf{r}},t).\,{\bf A}({\bf{r}},t)\nonumber \\
&-& \frac{1}{2} \int_0^\infty  \mbox{d} \omega\,
\left({{\bar{\bar{g}}_{e}}({\bf{r}},\omega ) \cdot {\bf{A}}
({\bf{r}},t)}\right)^{2}+ {W_{coul}}.
\end{eqnarray}
By using the Hamiltonian density~(\ref{Hamiltonian}) and recalling the
commutation relations~(\ref{commutation relation for A})
and~(\ref{commutation relation for XYp}), it is straightforward to
prove that the Heisenberg equations for the vector potential, the
transverse electric field and the particle coordinates yield the correct Maxwell equations and the Newtonian equation of
motion in the quantum domain. Let us begin with the Heisenberg equations for the vector potential
and the transverse electric field. Thus, the time derivative of~${\bf A}$ and~${\bf E}^{\bot}$ in a straightforward manner is given by
\begin{subequations}\label{Heisenberg eq of A & E}
\begin{eqnarray}
\dot{{\bf A}}({\bf r} , t )&=&\frac{1}{i \hbar}\left[ {\bf A}({\bf r} , t ) , {\cal H}\right]= - {\bf E}^{\bot}({\bf r} , t ), \\
\varepsilon_{0}\dot{{\bf E}}^{\bot}({\bf r} , t ) &=& \frac{1}{i
\hbar}\left[ \varepsilon_{0}{{\bf E}}^{\bot}({\bf r} , t ) , {\cal
H}\right]=\frac{{\bf \nabla}\times {\bf \nabla}\times {\bf A}({\bf
r} , t )}{\mu_{0}}
\nonumber \\
&-&{\bf \nabla} \times {\bf M}({\bf r} , t
)-\dot{\bf{P}}^{\bot}({\bf r} , t )-{\bf J}^{\bot}({\bf r} , t )\label{Heisenberg eq of E}
\end{eqnarray}
\end{subequations}
By using the constitutive equations of the displacement field ${\bf
D}^{\bot} = \varepsilon_{0}{\bf E}^{\bot} +{\bf{P}}^{\bot} $ and the
magnetic field strength ${\bf H}= {\bf B}/{\mu _{0}} - {\bf M}$,
Eqs.(\ref{Heisenberg eq of A & E}) lead to $\dot{{\bf
D}}^{\bot}({\bf r} , t ) = {\bf \nabla}\times {\bf H}^{\bot}({\bf r}
, t )- {\bf J}^{\bot}({\bf r } , t )$ and $ \dot{{\bf B}}({\bf r } ,
t ) = - {\bf \nabla}\times {\bf E}({\bf r } , t ) $ as expected,
where ${\bf D}^{\bot}$ is the transverse displacement field and
${\bf J}^{\bot}$ is transverse component of current density. In the
presence of charged particles, the longitudinal components of
electric and displacement fields can be written respectively as
\begin{subequations}\label{longitudinal component}
\begin{eqnarray}
{\bf E}^{\parallel}({\bf r } , t ) &=& -\frac{{\bf P}^{\parallel}({\bf r } , t )}{\varepsilon_{0}}- \nabla \varphi_{A},\label{longitudinal component
E}
\\
{\bf D}^{\parallel}({\bf r } , t ) &=&\varepsilon_{0}{\bf
E}^{\parallel}({\bf r } , t )+{\bf P}^{\parallel}({\bf r } , t )=
-\varepsilon_{0}\nabla \varphi_{A}\label{longitudinal component D}.
\end{eqnarray}
\end{subequations}
The Heisenberg equation of the charged particles in presence of~${\bf E}$ and~${\bf B}$ leads to the quantum mechanical version of Lorentz
force, namely,
\begin{eqnarray}\label{Heisenberg eq of charged particles}
m\ddot{\bf r}_{\alpha}= \frac{1}{i \hbar}\left[ m\dot{r}_{\alpha} , \cal{H}\right] = e_{\alpha}{\bf E}({\bf r}_{\alpha} , t ) + e_{\alpha}\dot{\bf
r}_{\alpha}\times {\bf B}({\bf r}_{\alpha} , t ).\nonumber \\
\end{eqnarray}
Calculations analogous to those of~(\ref{Heisenberg eq of A & E})
give the following Heisenberg equations for the dynamical variables
$ {\bf X}_{\omega} $ and $ {\bf Y}_{\omega} $, respectively, as
\begin{subequations}\label{Heisenberg eq of XY}
\begin{eqnarray}
\ddot{{\bf X}}_{\omega} ({\bf r} , t )& =&  -\omega^{2}{\bf X}_{\omega} ({\bf r} , t ) +{\bar{ \bar g}}_{e} ({\bf r} , t )\cdot{\bf E}({\bf r} , t
),\label{Heisenberg eq of X}\\
\ddot{{\bf Y}}_{\omega} ({\bf r} , t ) &=& -\omega^{2}{\bf
Y}_{\omega} ({\bf r} , t ) +{\bar{ \bar g}}_{m} ({\bf r} , t
)\cdot{\bf B}({\bf r} , t )\label{Heisenberg eq of Y}.
\end{eqnarray}
\end{subequations}
The formal solution of Eq.(\ref{Heisenberg eq of X}) is obtained as
\begin{eqnarray}\label{solution of X}
{\bf X}_{\omega} ({\bf r} , t ) &=& \left(\dot{{\bf X}}_{\omega} ({\bf r} ,0 ) \frac{\sin \omega t}{\omega} +  {\bf X}_{\omega} ({\bf r} , 0 ) \cos
\omega t \right) \nonumber \\
&+&{\bar{\bar g }}_{e} ({\bf r} , \omega ) \cdot \int_0^t d t^{\prime}
\,{\bf E}({\bf r} , t^{\prime}) \frac{\sin \omega( t -
t^{\prime})}{\omega}.
\end{eqnarray}
A similar relation also holds for $ {\bf Y}_{\omega} ({\bf r} , t )
$. To facilitate the calculations, let us introduce the following
annihilation operators:
\begin{subequations}\label{annihilation operators}
\begin{eqnarray}
{\bf f}_{e}({\bf r},\omega , t) = \frac{1}{\sqrt{2 \hbar \omega}} \left[-i \omega {\bf X}_{\omega } ({\bf r} , t ) + {\bf Q}_{\omega } ({\bf r} , t
)\right] , \label{annihilation operatorsfe} \\
{\bf f}_{m}({\bf r},\omega , t) = \frac{1}{\sqrt{2 \hbar \omega}}
\left[ \omega {\bf Y}_{\omega } ({\bf r} , t ) + i {\bf \Pi}_{\omega
} ({\bf r} , t )\right] ,\label{annihilation operators fm}
\end{eqnarray}
\end{subequations}
where ${\bf f}_{e}$ and ${\bf f}_{m}$ denote two independent
infinite sets of bosonic operators, which associated with the electric and
magnetic excitations of the system. By making use of Eqs.~(\ref{commutation relation for X})
and~(\ref{commutation relation for Y}), it is easily seen that the
bosonic operators have the commutation relations of the form
\begin{subequations}\label{commutation relations of operators}
\begin{eqnarray}
&&\hspace{-0.7cm}\left[{\rm f}_{ej}({\bf r},\omega , t) , {\rm f}^{\dagger}_{e
{j^{\prime}}}({\bf r}^{\prime},\omega^{\prime} , t) \right] = \delta
_{jj^{\prime} }\delta(\omega - \omega ^{\prime})\delta({\bf r} -{\bf
r}^{\prime} ), \\
&&\hspace{-0.7cm}\left[ {\rm f}_{mj}({\bf r},\omega , t) , {\rm
f}^{\dagger}_{m j^{\prime}}({\bf r}^{\prime},\omega^{\prime} , t)
\right] = \delta _{jj^{\prime} }\delta(\omega - \omega
^{\prime})\delta({\bf r} -{\bf
r}^{\prime}).\,
\end{eqnarray}
\end{subequations}
We can now invert Eq.~(\ref{annihilation operators}) to obtain the material
field  ${\bf X}_{\omega} $ and $ {\bf Y}_{\omega} $ in term of the
bosonic operators ${\bf f}_{e}$ and ${\bf f}_{m}$. With this in
mind, the polarization and magnetization fields of the
medium~(\ref{P and M field}) in term of the bosonic operators are
written as
\begin{subequations}\label{polarization&magnetization field}
\begin{eqnarray}
&&{\bf P} ({\bf r} , t ) = \varepsilon_{0}\int_ {0}^{\infty} \mbox{d} t^{\prime}\,\bar{ \bar{\chi}}_{e}({\bf  r} , t -t^{\prime} ) \cdot {\bf E} ({\bf r} ,
t^{\prime} ) + {\bf P}^{N} ({\bf r} , t ),  \nonumber \\ \\
&&{\bf M} ({\bf r} , t ) = \mu _{0}^{-1}\int_ {0}^{\infty} \mbox{d}
t^{\prime}\,\bar{ \bar{\chi}}_{m}({\bf  r} , t -t^{\prime} ) \cdot
{\bf B} ({\bf r} , t^{\prime} ) + {\bf M}^{N} ({\bf r} , t ),\nonumber \\
\end{eqnarray}
\end{subequations}
where electric and magnetic susceptibilities tensors of the medium are respectively defined as
\begin{subequations}\label{susceptibility}
\begin{eqnarray}
\hspace{-1cm}\bar{ \bar{{\boldsymbol \chi}}}_{e}({\bf  r} , t )=\Theta (t)\,{\varepsilon} _{0}^{-1}\int_ {0}^{\infty} \mbox{d} \omega\, {\bar{\bar{\boldsymbol
g}}}_{e}^{t}\cdot {\bar{\bar {\boldsymbol g}}}_{e} ({\bf  r} , \omega) \frac{\sin \omega t}{\omega}, \\
\hspace{-1cm}\bar{ \bar{{\boldsymbol \chi}}}_{m}({\bf  r} , t )= \Theta
(t)\,\mu_{0}\int_ {0}^{\infty} \mbox{d} \omega\,
{\bar{\bar{\boldsymbol g}}}_{m}^{t}\cdot {\bar{\bar {\boldsymbol g}}}_{m} ({\bf  r} , \omega) \frac{\sin
\omega t}{\omega}.
\end{eqnarray}
\end{subequations}
%
Here, the superscript $t$ indicates the transpose of a tensor.

Let $\bar{ \bar{{\boldsymbol \chi}}}_e({\bf  r} , \omega)$ and
$\bar{ \bar{{\boldsymbol \chi}}}_m({\bf  r} , \omega)$ ,
respectively, be the electric and the magnetic susceptibilities
tensors in frequency space. Then the electric permittivity and the
magnetic permeability tensors of the medium in term of the susceptibilities
tensors are written as $\bar{
\bar{{\boldsymbol \varepsilon}}}({\bf  r} , \omega)=\bar{
\bar{{\boldsymbol {\rm I}}}}+\bar{ \bar{{\boldsymbol \chi}}}_e({\bf
r} , \omega)$ and $\bar{ \bar{{\boldsymbol \mu}}}^{-1}({\bf  r} ,
\omega)=\bar{ \bar{{\boldsymbol {\rm I}}}}-\bar{ \bar{{\boldsymbol
\chi}}}_m({\bf  r} , \omega)$, where $\bar{ \bar{{\boldsymbol {\rm
I}}}}$ is the identity tensor. These are complex tensors of frequency which their real and
imaginary parts satisfy Kramers-Kronig relations and their dependence on coupling tensors, ${\bar{\bar{\boldsymbol {\rm
g}}}}_{e} $ and ${\bar{\bar{\boldsymbol {\rm g}}}}_{m}$, are given through the susceptibilities tensors as:
\begin{subequations}\label{susceptibility in frequency space}
\begin{eqnarray}
\bar{\bar{{\boldsymbol \chi}_{e}}}({\bf  r} , \omega ) &=&\varepsilon _{0}^{-1} \int_ {0}^{\infty} \mbox{d} \omega ^{\prime}\,\frac {
{\bar{\bar{\boldsymbol g}}}_{e}^{t}\cdot {\bar{\bar {\boldsymbol g}}}_{e} ({\bf  r} , \omega') } {\omega^{\prime 2} - \omega ^{2}+ i 0^{+}}, \\
{\bar {\bar{ {\boldsymbol \chi} } } _m}({\bf{r}},\omega)&=& \mu_{0}
\int_ {0}^{\infty} \mbox{d} \omega ^{\prime}\,\frac {{\bar{\bar{\boldsymbol g}}}_{m}^{t}\cdot {\bar{\bar {\boldsymbol g}}}_{m} ({\bf  r} , \omega') }
{\omega^{\prime 2} -
\omega ^{2}+ i 0^{+}}.
\end{eqnarray}
\end{subequations}
Given the electric permittivity and the magnetic permeability tensors of medium
, we can inverse the relations~(\ref{susceptibility in
frequency space}) and obtain the coupling tensors in term of these response tensors. Therefore, we find
%
\begin{subequations}\label{tensor g_e and g_m}
\begin{eqnarray}
{\bar{\bar{\boldsymbol g}}}_{e}^{t}\cdot {\bar{\bar {\boldsymbol g}}}_{e} ({\bf  r} , \omega)&=&
\frac{2{\varepsilon} _{0}\omega}{\pi} {\rm Im}\left[ {\bar{\bar
{\boldsymbol \varepsilon}}}({\bf  r} , \omega)\right],\label{tensor g_e}\\
{\bar{\bar{\boldsymbol g}}}_{m}^{t}\cdot {\bar{\bar {\boldsymbol g}}}_{m} ({\bf  r} , \omega)&=&
-\frac{2\omega}{\pi\mu_{0}} {\rm Im}\left[ {\bar {\bar {\boldsymbol
\mu}}}^{-1}({\bf r} , \omega)\right]. \label{tensor g_m}
\end{eqnarray}
\end{subequations}
The fields $ {\bf P}^{N} $ and $ {\bf M}^{N} $ in
Eqs.~(\ref{polarization&magnetization field}) are, respectively, the
noise polarization and the
noise magnetization operators which associated to the dissipation effects within medium.
As in the phenomenological method, we can separate the
positive and negative parts of fields, like $ {\bf P}^{N} =  {\bf
P}^{N(+)} + {\bf P}^{N(-)} $, where ${\bf P}^{N(+)} $ is the
conjugate of the negative part (analogously for ${\bf M}^{N(+)}
$) and yields
\begin{subequations}\label{e & m polarization noise}
\begin{eqnarray}
{\bf P}^{N(+)}({\bf  r} , t ) &=& i \int_ {0}^{\infty} \mbox{d}\omega \sqrt{\frac{ \hbar}{2\omega}}\, {\bar{\bar g}}_{e} ({\bf  r} , \omega) \cdot {\bf
f}_{e}({\bf r},\omega , 0)\, e^{-i\omega t }, \nonumber \\ \\
{\bf M}^{N(+)}({\bf  r} , t ) &=&  \int_ {0}^{\infty} \mbox{d}\omega
\sqrt{\frac{ \hbar}{2\omega}}\, {\bar{\bar g}}_{m} ({\bf  r} ,
\omega) \cdot {\bf f}_{m}({\bf r},\omega , 0)\, e^{-i\omega t },\nonumber \\
\end{eqnarray}
\end{subequations}
By taking the time derivative of Maxwell's equations
Eq.~(\ref{Heisenberg eq of E}) and using
Eq.~(\ref{polarization&magnetization field}), we obtain the
frequency-domain wave equation for the positive-frequency part of
the vector potential as,
\begin{eqnarray}\label{E^{+}}
\hspace{-1cm}{\nabla} \times {\bar{\bar {\mu}}}^{-1}{ \nabla} \times {\bf E}^{(+)}({\bf  r} , \omega) - \frac{{\omega}^{2}}{c^{2}}\, \bar{\bar \varepsilon}({\bf
r} , \omega) \,{\bf E}^{(+)}({\bf  r} , \omega) \nonumber \\
= \mu_{0}\omega^{2}\,{\bf P}^{N(+)}({\bf  r} , \omega ) + i\mu_{0}\omega
{\bf \nabla} \times {\bf M}^{N(+)}({\bf  r} , \omega ).
\end{eqnarray}
The formal solution of the above equation may be obtained through
finding an appropriate Green tensor. We thereby arrive at the following
expression for the electric field
\begin{eqnarray}\label{solotion of E^{+}}
\hspace{-1cm}{\bf E}^{(+)}({\bf  r} , t ) = ( i \omega{\mu_{0}})\int_ {0}^{\infty} \mbox{d}\omega \int \mbox{d}^{3}{ r}^{\prime}\, \bar{\bar {\boldsymbol {\rm G}}}({\bf  r} ,
{\bf  r}^{\prime},\omega) \cdot  \nonumber \\
\hspace{-1cm}\left[ -i\omega {\bf P}^{N(+)}({\bf  r}^{\prime} , \omega ) + {\bf
\nabla} \times {\bf M}^{N(+)}({\bf  r}^{\prime} , \omega
)\right] e^{-i\omega t },
\end{eqnarray}
where $ \bar{\bar G}({\bf  r} , {\bf  r}^{\prime},\omega) $ is the
classical Green tensor that satisfying the inhomogeneous Helmholtz equation with the space- and frequency-dependent complex permittivity and permeability of medium,
\begin{eqnarray}\label{classical Green tensor}
\hspace{-2cm}&&{\bf \nabla} \times \left[ \bar{\bar \mu}^{-1}({\bf  r} , \omega ) {\bf \nabla} \times \bar{\bar G}({\bf  r} , {\bf
r}^{\prime},\omega)\right]- \nonumber \\
&&\frac{\omega^{2}\,\bar{\bar \varepsilon}({\bf  r} ,
\omega)}{c^{2}}\,\,\bar{\bar G}({\bf  r} , {\bf  r}^{\prime},\omega)
= \delta ^{3} ({\bf  r} - {\bf  r}^{\prime})\bar{\bar{{\bf I }}},
\end{eqnarray}
The set of Eqs.~(\ref{e & m polarization noise}),~(\ref{E^{+}}),
and~(\ref{solotion of E^{+}}) together with the commutation
relations~(\ref{Heisenberg eq of A & E}), provide us with the electromagnetic field quantization in an anisotropic, dissipative and dispersive
magnetodielectric medium. It is easily seen that these relations are the same relations which were obtained via the
phenomenological quantization method~\cite{Matloob 1995}-~\cite{Dung 2003}. Thus, based on a rigorous quantization scheme, we arrive at the identical results.
\begin{figure}[ht]
\begin{minipage}[b]{0.5\linewidth}
\centering
\includegraphics[width=\textwidth]{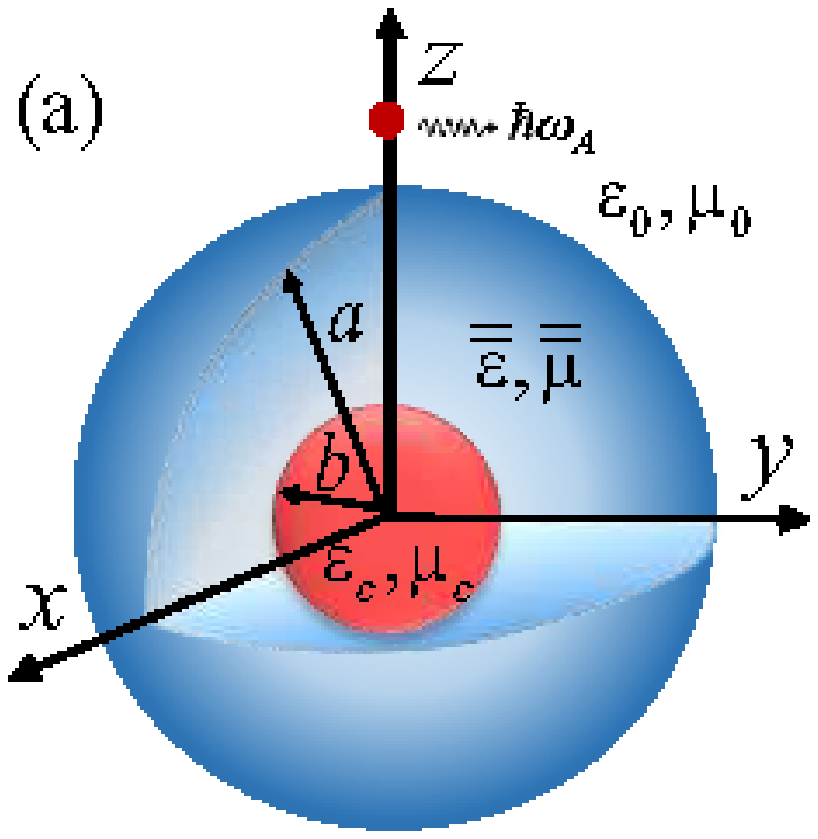}
\end{minipage}
\hspace{1.2cm}
\begin{minipage}[b]{0.55\linewidth}
\centering
\includegraphics[width=\textwidth]{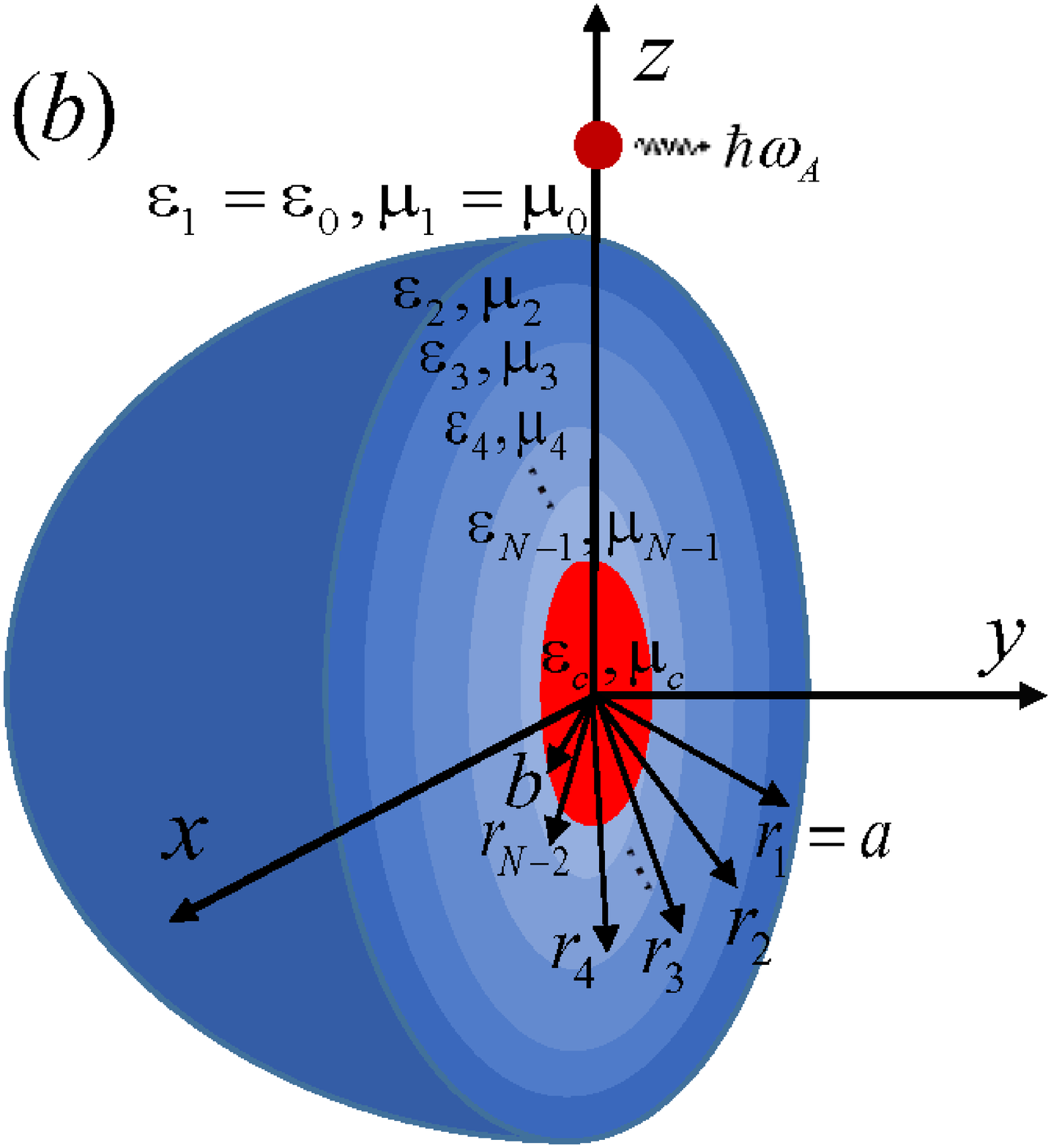}
\end{minipage}
\caption{ (a) The schematic of a spherical invisibility cloak with the
distribution of the material parameters which are given by Eq.~(\ref{permittivity & permeability _t & r}).
The clock shell has inner and outer radius of $b$ and $a$, respectively, and in the central region, $r < b$, is a homogeneous and isotropic material with the electric primitivity and the magnetic permeability function $\varepsilon_c$ and  $\mu_c$ as a object should be hidden. The atom is placed at the distance $r$ from the center of the cloak. Here, the media outside the cloak is free space. (b) The spherical invisibility cloak in part (a) is modeled by a large number of concentric layers with identical thickness. } \label{Fig:model}
\end{figure}
%

\section{spontaneous decay of an excited two-level atom }
\subsection{The model}
%
Let us consider an excited two-level atom with transition frequency $\omega_{\rm A}$ and the dipole
moment $d_{\rm A}$ placed at the point ${\bf r}_{\rm A}$ in the vacuum near an ideal invisibility cloaking. To simplify the treatment of the problem, we
consider a spherical hidden object which is covered by an invisibility spherical shell. Furthermore, with regard to the symmetry of this cloak, we
assume that the atom located on $z$ axis. Our approach can be simply extended to other type of cloaks and other locations. In this case, the cloak in
the annular region $a <r<b$ is a kind of rotationally uniaxial media characterized by~\cite{Pendry 2006}
\begin{eqnarray}\label{permittivity}
\bar{\bar\varepsilon } \left( {\bf r,\omega } \right) =
\left[ {\left( {{\varepsilon _r} - {\varepsilon _t}} \right)}
\right]\hat r\hat r + {\varepsilon _t}\bar{\bar I},
\end{eqnarray}
\begin{eqnarray}\label{permeability}
\bar{\bar \mu } \left( {\bf r,\omega } \right) = \left[
{\left( {{\mu _r} - {\mu _t}} \right)} \right]\hat r\hat r + {\mu
_t}\bar{\bar I},
\end{eqnarray}
where $ {\bar{\bar I}} = \hat r\hat r + \hat \theta \hat \theta  + \hat \varphi \hat \varphi \ $ is the unit dyad, the subscripts $r$
and $t$ denote the parameters along radial $\hat{r}$ and tangential
direction $\hat{\theta}$ or $\hat{\phi}$, respectively, and the permittivity
and permeability tensor components for the cloak shell are given by
\begin{subequations}\label{permittivity & permeability _t & r}
\begin{eqnarray}
\hspace{-0.5cm}{\varepsilon _t}\left( \omega  \right) &=& {\mu _t}\left( \omega
\right) = \frac{{{b}}}{{{b} - {a}}}\kappa_L(\omega),\label{permittivity & permeability _t} \\
\hspace{-0.5cm}{\varepsilon _r}\left( {\bf r,\omega } \right) &=& {\mu _r}\left(
{\bf r,\omega } \right) = \frac{{{b}}}{{{b} - {a}}}\left(
{\frac{{{{\left( {r - {a}} \right)}^2}}}{{{r^2}}}}\label{permittivity & permeability_r}
\right)\kappa_L(\omega).
\end{eqnarray}
\end{subequations}
Here, $b$ is the inner radius (radius of hidden object) and $a$ is the outer radius of the cloaking shell [see Fig.~\ref{Fig:model}(a)].
As mentioned before in the introduction, to achieve above material parameters in experimental, the most of the cloak device are constructed with
metamaterials consisting of resonating structures. This structure inevitably shows high loss and dispersion. For this purpose, we adopt a
single-resonance Lorentz
models for both the permittivity and permeability~\cite{Jackson 1999}. Thus, the permittivity
and permeability tensor components of the cloak have been
multiplied by a lorentzian factor, $\kappa_L(\omega) = \left( {1 +
\frac{\omega _p^{2}}{{\omega _0^2 - {\omega ^2} - i\gamma \omega }}}
\right)$, to consider the material absorbtion and dispersion of metamaterial structures.
Here, $ {\omega _p} $ and $ {\omega _0} $ are respectively the plasma
frequency and the resonant frequency and $ {\gamma} $ is the absorption coefficient of the cloaking. Without
loss of generality, we assume a homogenous and
isotropic object as a hidden object placed in the central region $ r < b $ with the material parameters $\varepsilon_{\rm c} =\mu_{\rm c} = \alpha \kappa_L(\omega)$ where
$\alpha $ is a constant.

The results obtained in the previous section can now be used to study the spontaneous emission of an excited two-level atom near such cloaked object. By inserting Eqs.~(\ref{e & m polarization noise}) and~(\ref{solotion of E^{+}}) into~(\ref{Hamiltonian}), the Hamiltonian of the whole system under the electric-dipole approximation and the
rotating wave approximation is recast to the following convenient form (refer to~\cite{Kheirandish 2010})
\begin{eqnarray}\label{Hamiltonian of cloak}
\hat H &=& \sum\limits_{\lambda  = e,m} {\int {{\mbox{d}^3}r\int_0^\infty
{\mbox{d}\omega \hbar \omega \,{\bf{\hat f}}^{\dag }_\lambda } } }
\left( {{\bf{r}},\omega } \right) \cdot {{{\bf{\hat f}}}_\lambda }\left( {{\bf{r}},\omega } \right)\nonumber \\
&+&\,\hbar {\omega _A}{{\hat \sigma }^\dag }\hat
\sigma  - \left[ {{{\hat \sigma }^\dag }{{\bf{d}}_A}\cdot
\int_0^\infty  {\mbox{d}\omega {\hat{\bf E}}^{(+)}\left( {{{\bf{r}}_A},\omega }
\right)} } + \rm{H}.c.\right],
\end{eqnarray}
Here, $\hat \sigma  = \left| l \right\rangle \left\langle u \right| $ and $
{{\hat \sigma }^\dag } = \left| u \right\rangle \left\langle l\right| $ are respectively the atomic lowering and raising operators where $\left| u \right\rangle $ ($ \left| l \right\rangle $) is the upper (lower) state of the atom whose energy is $ \hbar
{\omega _A} $(zero). Furthermore, $ {{\bf{d}}_A} $ is
transition dipole moment which defined as $ {{\bf{d}}_A} = \left\langle l \right|{\hat{\bf {d}}_A}\left| u \right\rangle  = \left\langle u \right|{\hat{\bf {d}}_A}\left| l \right\rangle $.
Since the spontaneous decay of an initially excited
atom is studied here, the state of the whole of system at time $ t $ can be expanded into the ground and excited states of the composite system
including electromagnetic field and the cloak, $\left| \{0\} \right\rangle$ and $\left| {{{\bf{1}}_\lambda }\left(
{{\bf{r}},\omega } \right)} \right\rangle$, and the unperturbed atomic states as
\begin{eqnarray}\label{wave function}
&&\left| {\psi \left( t \right)} \right\rangle = {C_u}\left( t \right){e^{ - i{{\tilde \omega }_A}t}}\left| \{0\} \right\rangle \left| u \right\rangle
\nonumber \\
&&+ \sum\limits_{\lambda  = e,m} {\int {{\mbox{d}^3}r} } \int_0^\infty
{\mbox{d}\omega \,\,{e^{ - i\omega t}}{{\bf C}_{\lambda l}}} \left(
{{\bf{r}},\omega ,t} \right) \cdot \left| {{{\bf{1}}_\lambda }\left(
{{\bf{r}},\omega } \right)}  \right\rangle \left| l \right\rangle,
\nonumber \\
\end{eqnarray}
The population probability amplitudes of the upper and lower states of the whole
system, ${C_u} $ and $ {C_{\lambda
l}} $, can be easily calculated from the Schr\"{o}dinger
equation. In the case of ${C_u}\left( t \right) $, by making use of the Green tensor integral relationship that was proven in~\cite{Knöll 2001,Dung 2003}, and inserting the
initial conditions  $ {C_u}( 0 )=1 $ and $ {C_{\lambda}}( {\bf r},\omega ,0 ) = 0 $, the
following time evolution is yield:
\begin{eqnarray}\label{c_dot _u}
\hspace{-0.5cm}{\dot C_u}\left( t \right) =  - i\delta \omega \,{C_u}\left( t
\right) + \int_0^t {\mbox{d}t'\,{{\rm K}} \left( {t - t'} \right)\,}{C_u}\left( {t'} \right),
\end{eqnarray}
where the kernel function ${ {\rm K}}\left( {t - t'} \right)$ is
determined by the Green tensor of the system in the position of the atom
\begin{eqnarray}\label{kernel tensor}
{ {\rm K}} \left( {t - t'} \right) &=& -\frac{1}{{\hbar \pi {\varepsilon _0}}}\int_0^\infty  {\mbox{d}\omega } \frac{{{\omega ^2}}}{{{c^2}}}{e^{ - i\left(
{\omega
- {{\tilde \omega }_A}}
\right)( {t - t'} )}} \nonumber \\
& \times & {{\bf d_A}} \cdot {\rm{ Im}}[
\bar{\bar G} ( {{\bf r}_A},{\bf d}_A,{\omega _A} )] \cdot \,{\bf
d}_{A} ,
\end{eqnarray}
in which ${{\tilde \omega }_A} = {\omega _A} - \delta \omega $
is the shifted transition frequency of the atom in the presence of
the cloak and $ \delta \omega $ is the Lamb
shift.
It is easily seen that the coupled integro-differential
equation~(\ref{c_dot _u}) cannot be solved analytically.
But, we can gain a physical insight into the spontaneous emission process of the atomic system near the cloak on base of the analytical solutions. So,
we concentrate our attention to the limiting cases of weak and strong atom-field coupling.

\subsection{Weak and strong coupling regimes}
%
Let us first consider the weak coupling where the atom is only slightly perturbed by the vacuum fields, then the
Markov approximation applies. In this regime, the coefficient  $ {C_u}\left( {t'} \right) $ in Eq.~(\ref{c_dot _u}) can be replaced by
$ {C_u}\left( t \right) $ and the time integral $ \int_0^t {\mbox{d}t'} {e^{ - i\left( {\omega  - {{\tilde \omega }_A}}
\right)\left( {t - t'} \right)}}$ in Eq.(\ref{kernel tensor}) can be approximated by the zeta
function
$ \xi({{\tilde \omega }_A} - \omega)  $
where $ \xi(x)=\pi \delta \left( { x } \right) + iP\left( {\frac{1}{{ x }}} \right) $.
By applying these approximations, the probability amplitude ${C_u}\left( {t'} \right) $ is written as the form
\begin{eqnarray}\label{c_u2}
{C_u}\left( t \right)=\exp \left[ {\left( { - \frac{1}{2}\Gamma  + i\delta \omega } \right)t} \right],
\end{eqnarray}
where the Lamb shift $\delta \omega \ $ and the decay rate $ \Gamma $
are given respectively as
\begin{subequations}\label{lamb shift and decay rate}
\begin{eqnarray}
\hspace{-0.8cm}\delta \omega & = &\frac{1}{{\hbar \pi {\varepsilon
_0}}}\,{\rm P}\int_0^\infty  {\mbox{d}\omega } \frac{{{\omega
^2}}}{{{c^2}}}\frac{{{{\bf d}_A} \cdot {\mathop{\rm Im}\nolimits}
{\bar{\bar G}} \left( {{{\bf r}_A},{{\bf r}_A},\omega } \right)
\cdot {{\bf d}_A}}}{{\omega  - {{\tilde \omega }_A}}}, \label{lamb shift} \\
\hspace{-0.8cm}\Gamma & =& \frac{{2{{\tilde \omega }^2}_A}}{{\hbar {\varepsilon
_0}{c^2}}}\,{{\bf d}_A} \cdot \mathop{\rm Im}\nolimits {\bar{\bar G}}
\left( {{{\bf{r}}_A},{{\bf{r}}_A},{\tilde \omega }_A } \right) \cdot {{\bf
d}_A}. \label{decay rate}
\end{eqnarray}
\end{subequations}
It is obviously seen that the effect of cloak and their properties
are contained in the Green tensor. Therefore, in order to calculate the spontaneous emission rate of the excited atom, we first need to compute the
Green tensor of the system.
\begin{figure}[ht]
\begin{minipage}[b]{1\linewidth}
\centering
\includegraphics[width=\textwidth]{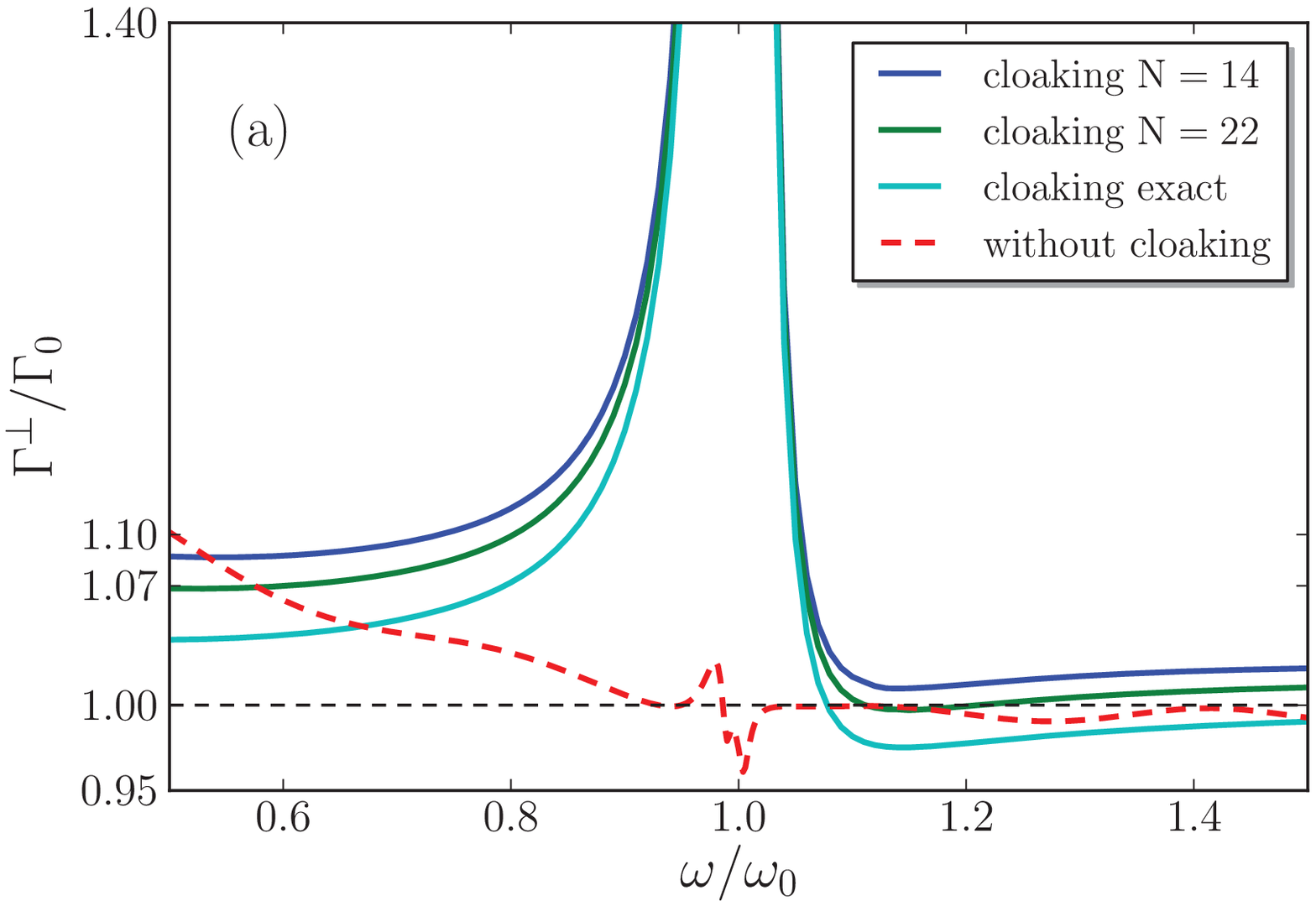}
\end{minipage}
\hspace{1.2cm}
\begin{minipage}[b]{0.95\linewidth}
\centering
\includegraphics[width=\textwidth]{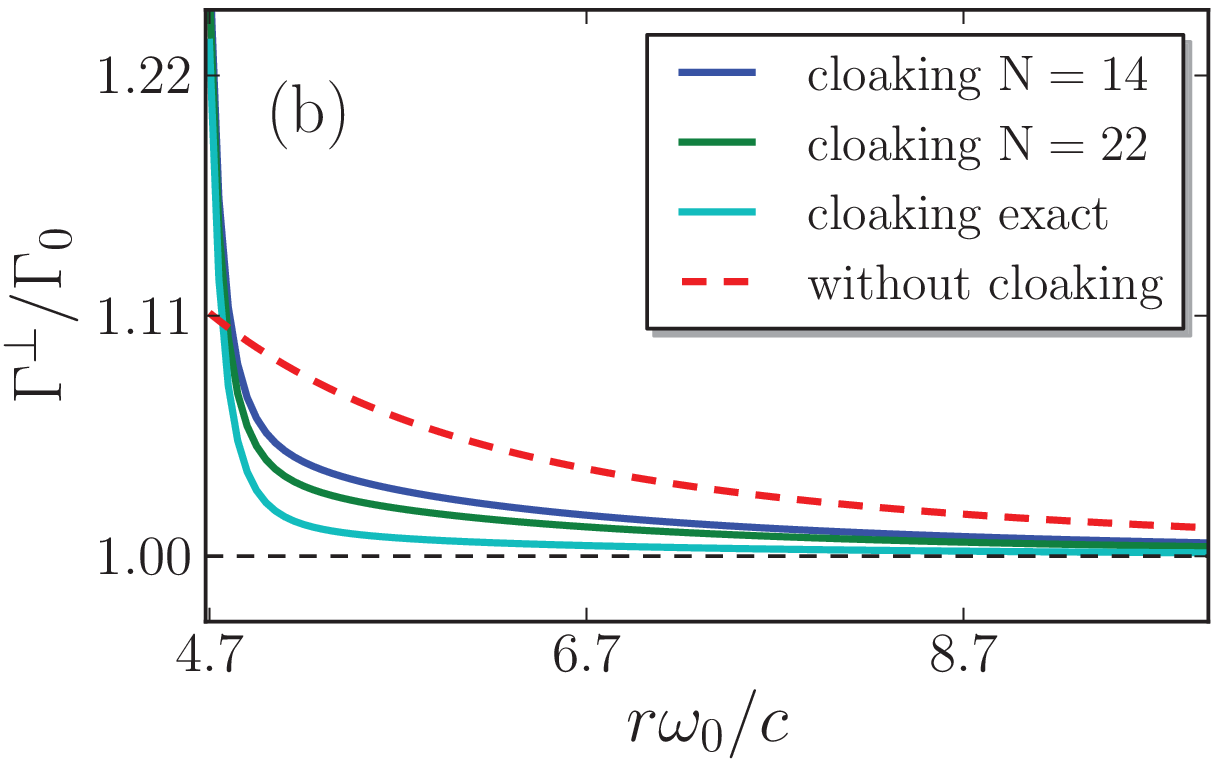}
\end{minipage}
\caption{ The vertical spontaneous emission rate (\ref{vertical
rate 1}) as a function of (a) the dimensionless frequency
$\omega_{A}/\omega_{0} $ and (b) the dimensionless distance
${r \omega_{0}}/{c} $.
The material absorbtion
and dispersion of the central hidden object and the cloak are described by the Lorentz model with parameters $ {\omega_{p} = 0.01\omega_{0}} $ and $\gamma = 0.01\omega_{0} $. The inner and outer radius of the cloak shell have been chosen $b= 3c/\omega_{0}$ and $a= 4.5c/\omega_{0}$, respectively, and the excited two-level atom placed at $ r_{A}=4.7c/\omega_{0}$.
For comparison, the spontaneous decay in free space is shown with black dashed line. } \label{Fig:layered}
\end{figure}

As seen from Eq.~(\ref{permittivity & permeability _t & r}), the cloak shell requires the material parameters with radius-dependent and anisotropic
characteristics. At the first glance, it seems that the calculation of the Green tensor of such system, which is needed for substitution in
Eq.~(\ref{decay rate}), is impossible. However, in the special case in which the cloak has a symmetric geometry shape such as the spherical cloak,
the evaluation of the Green tensor, though lengthy, is rather straightforward. We call this Green tensor extraction \textit{exact} method against another one which is approximately called \textit{discrete} method. The latter one is important at
least from one aspect: It offers the possibility of realizing such cloak by layered structures in experimental~\cite{Schurig 2006}, since the evaluation of the Green tensor is done by a discrete model of layered structure.

Similar to the practical realization of an anisotropic and inhomogeneous cylindrical cloak by concentric layered structure consisting of anisotropic
structure~\cite{Schurig 2006}, we can imagine a layered structure of \textit{homogeneous} and \textit{anisotropic} materials to mimic the ideal cloak by enforcing
the tangential components of the permittivity and the permeability of different layers to vary with the radius according to Eq.~(\ref{permittivity &
permeability _t & r}). The cloak that is modeled in this manner~[see Fig.\ref{Fig:model}(b)], is an example of a spherically layered magnetodielectric
medium for which its dyadic Green function is known~\cite{Tai 1994,Qiu 2007}. The details of these calculation are omitted here for the sake of brevity.
The
complete descriptions of both methods are given in Appendix~\ref{Appen:Green tensor}.

Of course, a cloaking structure was proposed in~\cite{Huang 2007, Qiu 2009}, that does not require metamaterials to realize the anisotropy or inhomogeneity of the material parameters. This allows us to realize the cloak through natural materials by using a layered structure of alternating \textit{homogeneous} and \textit{isotropic} materials. We do not consider this proposed method here.
%
\begin{figure*}[ht]
\begin{minipage}[b]{0.495\linewidth}
\centering
\includegraphics[width=\textwidth]{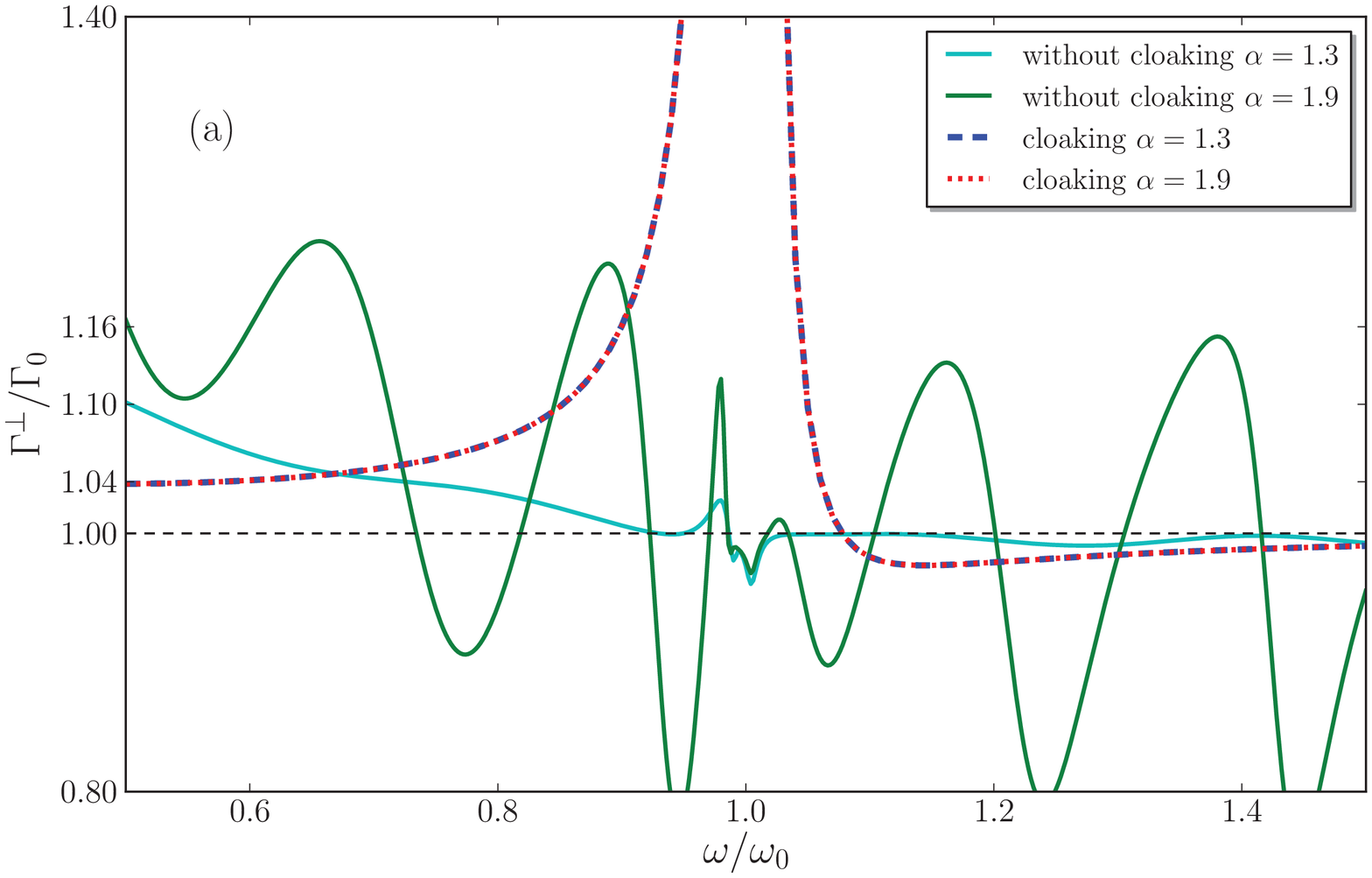}
\end{minipage}
\hspace{0 cm}
\begin{minipage}[b]{0.49\linewidth}
\centering
\includegraphics[width=8.4cm,height=5.5cm]{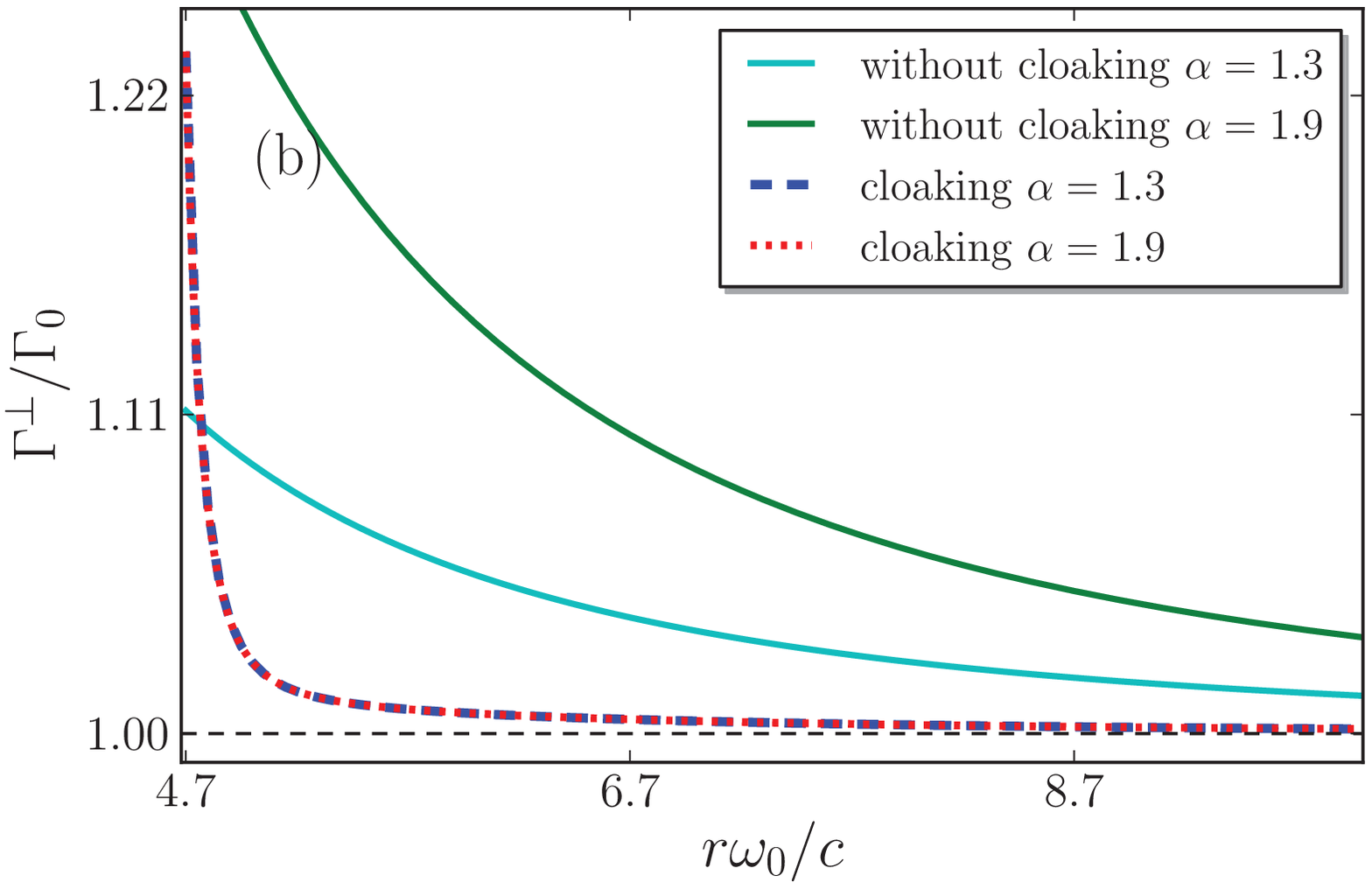}
\end{minipage}
\hspace{0 cm}
\begin{minipage}[b]{0.495\linewidth}
\centering
\includegraphics[width=\textwidth]{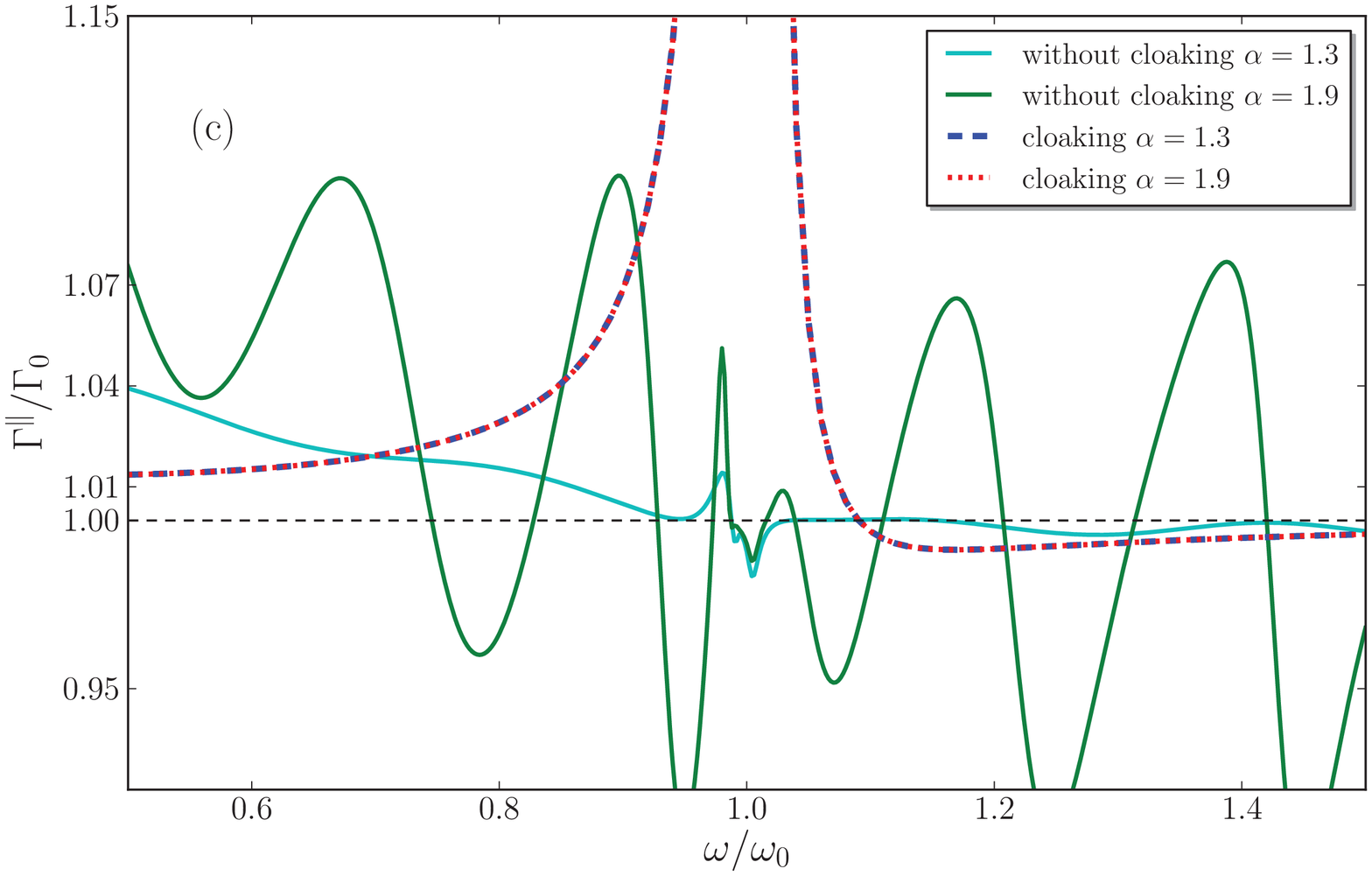}
\end{minipage}
\hspace{0 cm}
\begin{minipage}[b]{0.49\linewidth}
\centering
\includegraphics[width=8.5cm,height=5.5cm]{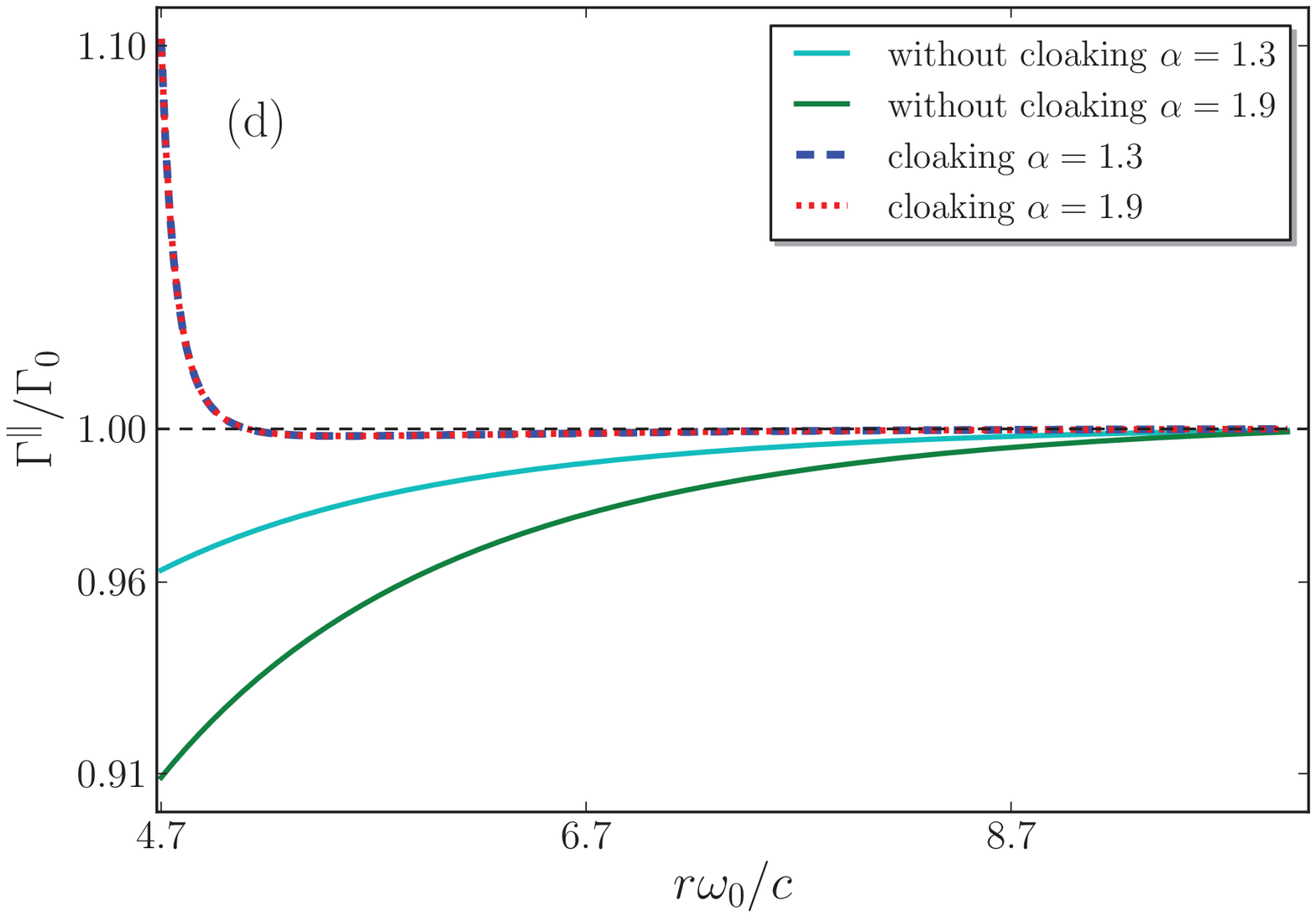}
\end{minipage}
\caption{ The vertical spontaneous emission rate (\ref{vertical rate 1}) as a function of (a) the dimensionless frequency
$\omega_{A}/\omega_{0} $ and (b) the dimensionless distance ${r \omega_{0}}/{c} $. Panels (c) and (d) are the same as panels (a) and (b) but for the case that the atomic dipole moment is parallel to the cloak shell and the spontaneous emission rate is calculated by Eq.~(\ref{tangential rate}).  Here, the parameters are identical to those used in Fig.~\ref{Fig:layered}.}
\label{Fig:object}
\end{figure*}
%

Now, we assume that the imaginary part of the Green tensor in
the resonance region of atom-field coupling has a
Lorentzian shape with the central frequency $\omega_{\rm c}$
and the half width at half maximum $\delta \omega_{\rm c}$. Unlike to the weak coupling, the zeta
function no longer act as the $\delta$ function. In this case, the frequency integral in Eq.~(\ref{kernel tensor}) can be done by expanding the limit
of integration to $\pm \infty$ which leads to~\cite{Dung 2000}
\begin{eqnarray}\label{strong kernel tensor}
{ \bar{\bar K}} \left( {t - t'} \right)
& =& - \frac{1}{2}\Gamma \delta {\omega _c}{e^{ - i\left( {{\omega _c} - {\omega _A}} \right)\left( {t - t'} \right)}}{e^{ - \delta {\omega
_c}\left|{t
- t'} \right|}}.
\end{eqnarray}
By substituting the above expression into Eq.~(\ref{c_dot _u}) and making the differentiation
of both sides of the resulting equation over time, one arrives at a second order homogeneous differential equation
\begin{eqnarray}\label{C dotdot u}
{\ddot C_u}\left( t \right) + \left[ {i\left( {{\omega _{\rm c}} - {\omega _A}} \right) + \delta {\omega _{\rm c}}} \right]{\dot C_u}( t ) +
({\Omega}/{2})^{2}{C_u}( t ) = 0, \nonumber \\
\end{eqnarray}
where $ \Omega  = \sqrt {2 \Gamma\, \delta {\omega _{\rm c}}} $. This is in the form of the damped harmonic oscillator equation with the formal
solution given for $\omega_A\approx\omega_c $ by
\begin{eqnarray}\label{general solution}
{C_u}(t)& = &\frac{1}{2}\left( {1 + \frac{{\delta {\omega _{\rm c}}}}{{\sqrt {\delta {\omega _{\rm c}}^2 - {\Omega ^2}} }}} \right){e^{\left( { -
\delta {\omega _{\rm c}} + \sqrt {\delta {\omega _{\rm c}}^2 - {\Omega ^2}} } \right)\frac{t}{2}}}\nonumber \\
&+& \frac{1}{2}\left( {1 - \frac{{\delta {\omega _{\rm c}}}}{{\sqrt {\delta {\omega _{\rm c}}^2 - {\Omega ^2}} }}} \right){e^{-\left( {\delta {\omega
_{\rm c}}+ \sqrt {\delta {\omega _{\rm c}}^2 - {\Omega ^2}}} \right)\frac{t}{2}}}.\nonumber \\
\end{eqnarray}
It is seen that when $\Omega\gg {\delta {\omega _{\rm c}}}$, the decay probability amplitude of the upper atomic state shows the well-known
phenomenon of damped Rabi oscillations
\begin{eqnarray}\label{c u strong}
{C_u}(t) = {e^{ - \frac{{\delta {\omega _m}}}{2}t}}\cos \left( {\Omega t/2} \right).
\end{eqnarray}
This is the principal signature of strong atom-field coupling. In the opposite case, when $\Omega\ \ll {\delta {\omega _{\rm c}}}$, we recover the weak
coupling result~(\ref{c_u2}) which obtained within the Markovian approximation.
%
\subsection{Analytical and numerical results}
%
Equation~(\ref{decay rate}) can now be applied to the atom which
its atomic dipole moment may be perpendicular to the interface
along the z axis, and/or parallel to the interface along the y
axis. We denote the former case by the superscript $\bot$ and the latter one by the superscript $\|$.
With the Green's tensor in hand given in [Appendix~\ref{Appen:Green tensor}] and use of the symmetry of our system, the scattering part of the Green tensor for the two special cases of radial and tangential direction are simplified as follows:
\begin{subequations}\label{vertical and tangential Green}
\begin{eqnarray}
{\bar{ \bar G}}_s^{(11)}\left( {{\bf r_A},{\bf r_A},\omega } \right)
&=& \frac{{i{k_1}{\mu _1}}}{{4\pi }}\sum\limits_{n = 0}^\infty  {n(n +
1)(2n + 1)}\nonumber\\
&\times& B_N^{11}{{\left( {\frac{{z_n ^{\left( 1
\right)}\left( {kr} \right)}}{{kr}}} \right)}^2},\label{vertical Green}\\
{\bar{ \bar G}}_s^{(11)}\left( {{{\bf r}_A},{{\bf r}_A},\omega } \right) &=& \frac{{i{k_1}{\mu _1}}}{{8\pi }}\sum\limits_{n = 0}^\infty  {(2n +
1)B_M^{11}{{\left( {z_n^{\left( 1 \right)}\left( {kr} \right)} \right)}^2}}\nonumber \\
&+& B_N^{11} {\left( {\frac{1}{{kr}}\frac{{\mbox{d}\left[ {rz_n^{\left( 1
\right)}\left( {kr} \right)} \right]}}{{kr}}} \right)^2}.\label{tangential Green}
\end{eqnarray}
\end{subequations}
It is worth noting that in the above derivation of Eq.~(\ref{vertical and tangential Green})
seams only one of the electromagnetic Green tensors of our system which have been obtained in Eqs.~(\ref{Green11}) and~(\ref{GsT})
to be used here. In fact, the exact Green tensor~(\ref{Green11})
has exactly the same form as the discrete Green tensor~(\ref{GsT})
for $f=1$ with one exception for the coefficients $B_{N,M}^{11}$ which given by Eqs.~(\ref{B_MN}) and~(\ref{B_MN layer}), respectively.
This follows from the assumption that the field point and source point in our case are located out of the cloak.

By substituting the expressions above into Eq.~(\ref{decay rate}), we arrive at the following formulas for the spontaneous
decay rate of the aforementioned atom at arbitrary position
\begin{subequations}\label{vertical and tangential rate}
\begin{eqnarray}
\frac{{{\Gamma ^ \bot }}}{{{\Gamma _0}}} &=& 1 + \frac{{6\pi }}{\omega }{\rm{Im}}\left[ {\frac{{i{k_1}{\mu _1}}}{{4\pi }}\sum\limits_{n = 0}^\infty
{n(n + 1)(2n + 1)\,\,\,} } \right. \nonumber\\
&&\left. {\times B_N^{11}{{\left( {\frac{{z_n^{\left( 1 \right)}\left( {kr} \right)}}{{kr}}} \right)}^2}} \right], \label{vertical rate 1}\\
\frac{{{{\rm{\Gamma }}^\parallel }}}{{{{\rm{\Gamma }}_0}}}& =& 1 + \frac{{6\pi }}{\omega }{{\rm Im}} \left[ {\frac{{i{k_1}{\mu _1}}}{{8\pi
}}\sum\limits_{n = 0}^\infty  {(2n + 1)B_M^{11}{{\left( {z_n^{\left( 1 \right)}\left( {kr} \right)} \right)}^2}} } \right. \nonumber \\
&&\left. + B_N^{11} { {{\left( {\frac{1}{{kr}}\frac{{\mbox{d}\left[
{rz_n^{\left( 1 \right)}\left( {kr} \right)} \right]}}{{kr}}}
\right)}^2}} \right], \label{tangential rate}
\end{eqnarray}
\end{subequations}
where $ {\Gamma _0} = \frac{{{{\tilde \omega }^3}_A \mbox{d}_A^2}}{{3\hbar \pi {\varepsilon _0}{c^3}}} $ is the rate of spontaneous emission in free space.

Fig.~\ref{Fig:layered} illustrates the variation of the normalized rate of the radial spontaneous decay given by Eq.~(\ref{vertical and tangential rate}) for a radially oriented transition dipole moment as a function of the dimensionless parameters $ {\omega }/{\omega _0}$ and ${r{\omega _0}}/{c}$. According to the experimental data reported in~\cite{Brune 1994} for the spontaneous emission rate of a quantum dot in the microwave frequency region and with regards that the majority of the cloak devices have been constructed in this
frequency range, the material parameters of the constructed cloak in~\cite{Schurig 2006} are used here to study the performance of the cloak quantum mechanically.

In Fig.~\ref{Fig:layered}, the cyan curve represents the exact result in which the exact Green tensor~(\ref{Green11}) with the coefficients~(\ref{B_MN}) are inserted in~Eq.~(\ref{vertical rate 1}), whereas, the green and the blue curves correspond to the cases that the clock is approximately modeled by $16$ and $22$ spherical thin layers, respectively, and $ {{\rm{\Gamma }}^ \bot }/{{\rm{\Gamma }}_0} $ are calculated by making use of the coefficients~(\ref{B_MN layer}). The radial component of the material parameters of each layer are given in Table.~(\ref{Table}).
To examine the performance of the clock shell, the normalized spontaneous decay $ {{\rm{\Gamma }}^ \bot }/{{\rm{\Gamma }}_0} $ is also depicted in the absence of the cloak. The corresponding plot is indicated by the red dashed curve.
For comparison, the spontaneous decay rate in free space is shown by black dashed line.
%

From Fig.~\ref{Fig:layered} (a), we observe a significant enhancement of the decay rate in the vicinity of the cloak resonance frequency.
%
%
It reveals that the cloak near the resonance frequency not only conceals the object but also makes it more visible. In contrast, far from the cloak resonance frequency, the spontaneous emission of the atom differs from the free space value by about $0.1$. While, the variations are in the order of $0.01$ in the absence of the cloak. It seams that the performance of the cloak shell is quite well far from the cloak resonance frequency.
But, with decreasing distance between the atom and the cloak, near-field effects become important and the spontaneous emission reveals the strong enhancement, as seen in Fig.~\ref{Fig:layered}(b). It describes non-radiative decay, i.e., the energy transfer from the atom to the cloak~\cite{Dung 2001}, and causes the hidden object to appear. The enhancement decreases by increasing distance between the atom and the cloak and all curves asymptotically tend to the free space value when $r \rightarrow \infty$.

From the classically researches reported on the cloak devices, we expect that the combination of the cloak and the object had the properties of free
space when viewed externally. Here, if the cloak can hide the object successfully, the spontaneous decay rate should be unchanged as if there were nothing. Whereas, our results demonstrate that there is a strikingly difference with the free space value when the atom located in close distance to the cloak and also near the cloak resonance frequency. So, just far from the cloak resonance frequency and moderate distance, the object plus the cloak are rather invisible.

Comparing the cyan curve with the green and the blue curves in Fig.~\ref{Fig:layered} show clearly that as the layers are made thinner by
increasing the number of layers from 14 to 22, the agreement between the approximate results and the exact one increases. Since, the layered cloak with a large number of layers has nearly the same property as the equivalent cloak shell.
This means that we are able to use a concentric layered structure to realize a spherical cloak shell. However, by increasing the number of layers, calculating the electromagnetic Green tensor of the system becomes computationally time-consuming due to the numerically demands of determining the coefficients~(\ref{B_MN layer}). In following, we therefore restrict our attention only to the exact Green tensor Eq.(\ref{Green11}) together with the coefficients which are given by Eq.~(\ref{B_MN}).

The effect of different hidden objects covered by the cloak shell on the spontaneous emission rate is illustrated in Fig.~\ref{Fig:object}. The different objects enter to our calculations by attributing different values to the factor $\alpha$ of the material parameters. By comparing the green and the cyan curves in Figs.~\ref{Fig:object} (a) and (b) it is seen that with increasing the constant factor $\alpha $ from $1.3$ to $1.9$ in the absence of the cloak, changes in amplitude of the spontaneous decay rate severely enhance. While, in agreement with the classical results, the blue dashed and red dotted curves stay without any changes in the presence of the cloak.
Therefore, the cloak acts independent of the object to be cloaked.

For a tangentially oriented dipole moment, the normalized rate of the spontaneous decay given by Eq.~(\ref{tangential rate}) are plotted in Figs.~\ref{Fig:object} (c) and~\ref{Fig:object} (d) as a function of the dimensionless parameters ${\omega }/{\omega _0}$ and ${r{\omega _0}}/{c}$, respectively. From Fig.~\ref{Fig:object} (c), it is seen that the spontaneous decay rate, except a subtle decrease in amplitude, shows a similar behavior as in Fig.~\ref{Fig:object} (a).
For instance, at frequency ${\omega }/{\omega _0}=0.5 $, the normalized decay rate~(\ref{tangential rate}) for the objects with factors $ \alpha=1.3 $ and $\alpha=1.9 $ are in the order of $1.04$ and $1.07$ in the absence of the cloak, respectively, and order of $1.01$ in the presence of the cloak. While, these values for a radially oriented dipole moment are in the order of $1.01$, $1.16$ and $1.04$ respectively.

Fig.~\ref{Fig:object} (d) represents that the results with only slight reduction in amplitude are similar to Fig.~\ref{Fig:object} (a), provided that the object surrounded by the cloak shell. In the absence of cloak and close distances to the cloak, there is a noticeable difference between Figs.~\ref{Fig:object} (b) and~\ref{Fig:object} (d). Finally, the tangential spontaneous decay rate like the vertical spontaneous decay rate tends to the unit value very far away from the cloak.

As a conclusion of this section, we found that such cloak shells independent of the dipole moment orientation of our probe which is here an excited two level atom, operate quite well far from the cloak resonance frequency and moderate distance. Particularly, the ability of the cloak to conceal an object with larger material parameters is better than when the cloak is absent.
%
\section{Spatial distribution of the emitted-light intensity}
%
Let us finally examine the influence of the
the combination of the cloak shell and the hidden object on the emission pattern of
light emitted by the excited atom.
The intensity of the spontaneously emitted light registered
by a photodetector at position $r$ and time $t$ is given by~\cite{Dung 2000,Dung 2001}
\begin{eqnarray}\label{intensity}
{\bf{I}}\left( {{\bf r},t} \right) \equiv \left\langle {\psi \left(
t \right)} \right|\hat{\bf E}^{(-)}\left( {{\bf r},\omega } \right) \cdot
{\hat{\bf{ E}}^{(+)}}\left( {{\bf {r}},\omega}
\right)\left| {\psi \left( t \right)} \right\rangle.
\end{eqnarray}
By substituting Eqs.~(\ref{solotion of E^{+}}) and~(\ref{wave function}) into the above equation and after some algebraic calculations, we get
\begin{eqnarray}\label{intensity2}
I\left( {{\bf{r}},t} \right) &=& \left| {\frac{{k_A^2{{\bf d}_A}}}{{\pi {\varepsilon _0}}}\cdot\int_0^t {\mbox{d}t'} } \right. \nonumber \\
&&\hspace{-1cm}\times{\left. {\left[ {{C_u}\left( {t'} \right)\left. {\int_0^\infty  {\mbox{d}\omega \, {\rm Im}[{\bar{ \bar G}}({\bf{r}},{{\bf{r}}_A},\omega )]{e^{ - i\left(
{\omega  - {\omega _A}} \right)\left( {t - t'} \right)}}} } \right]} \right.} \right|^2}. \nonumber \\
\end{eqnarray}
\begin{figure*}[ht]
\begin{minipage}[b]{0.495\linewidth}
\centering
\includegraphics[width=7.5cm,height=7.5cm]{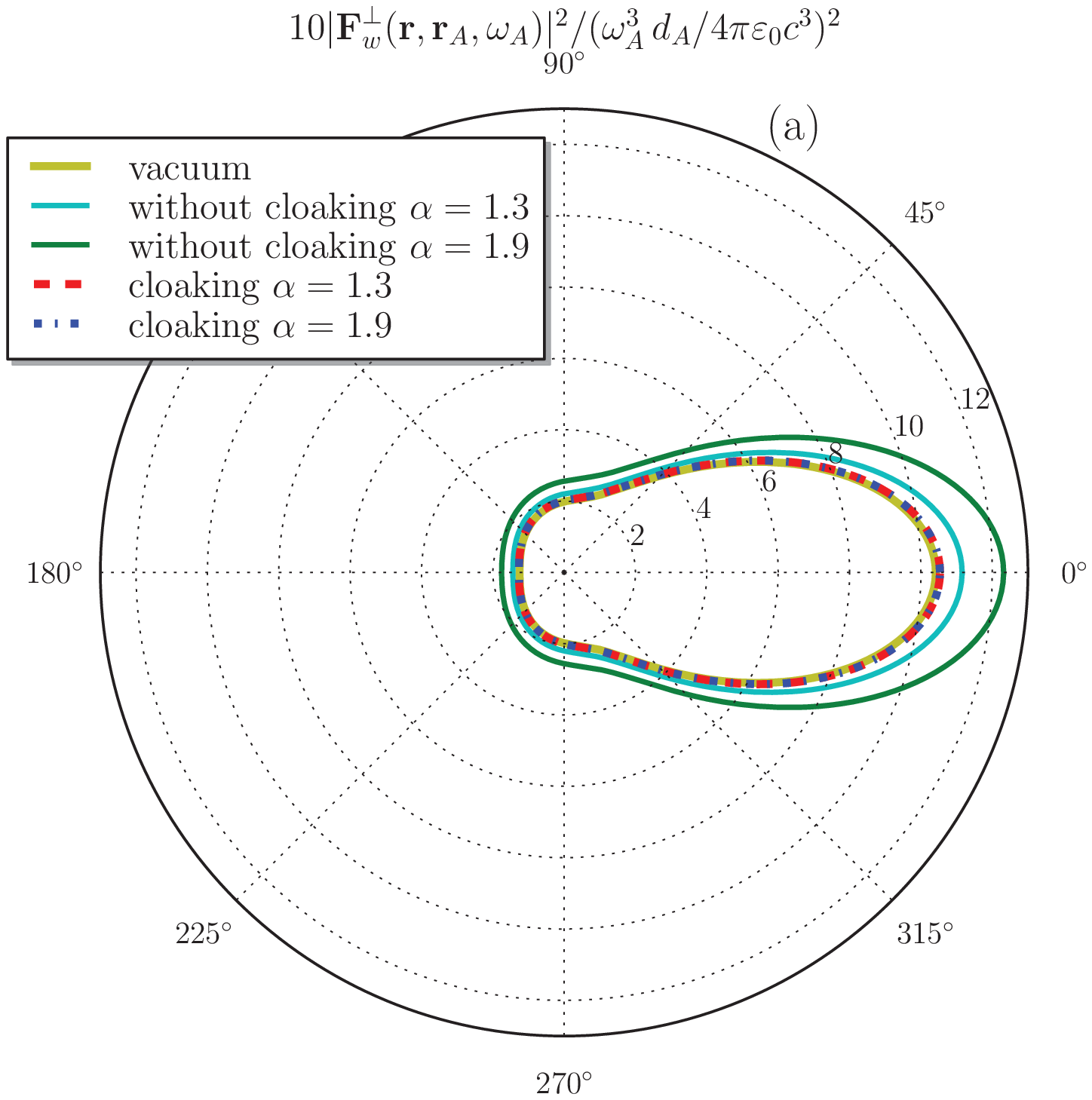}
\end{minipage}
\hspace{0 cm}
\begin{minipage}[b]{0.49\linewidth}
\centering
\includegraphics[width=7.5cm,height=7.5cm]{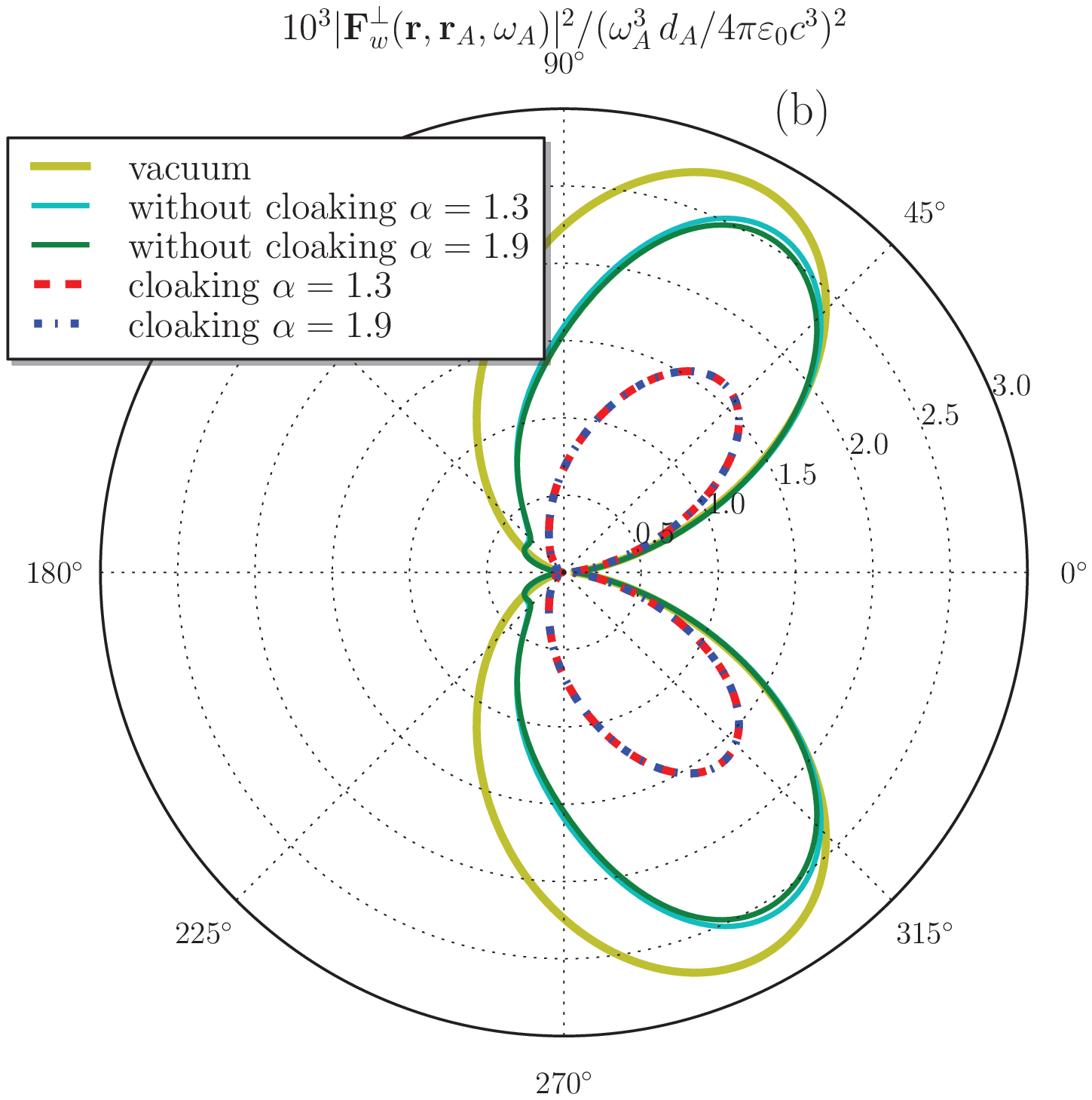}
\end{minipage}
\hspace{0 cm}
\begin{minipage}[b]{0.495\linewidth}
\centering
\includegraphics[width=7.5cm,height=7.5cm]{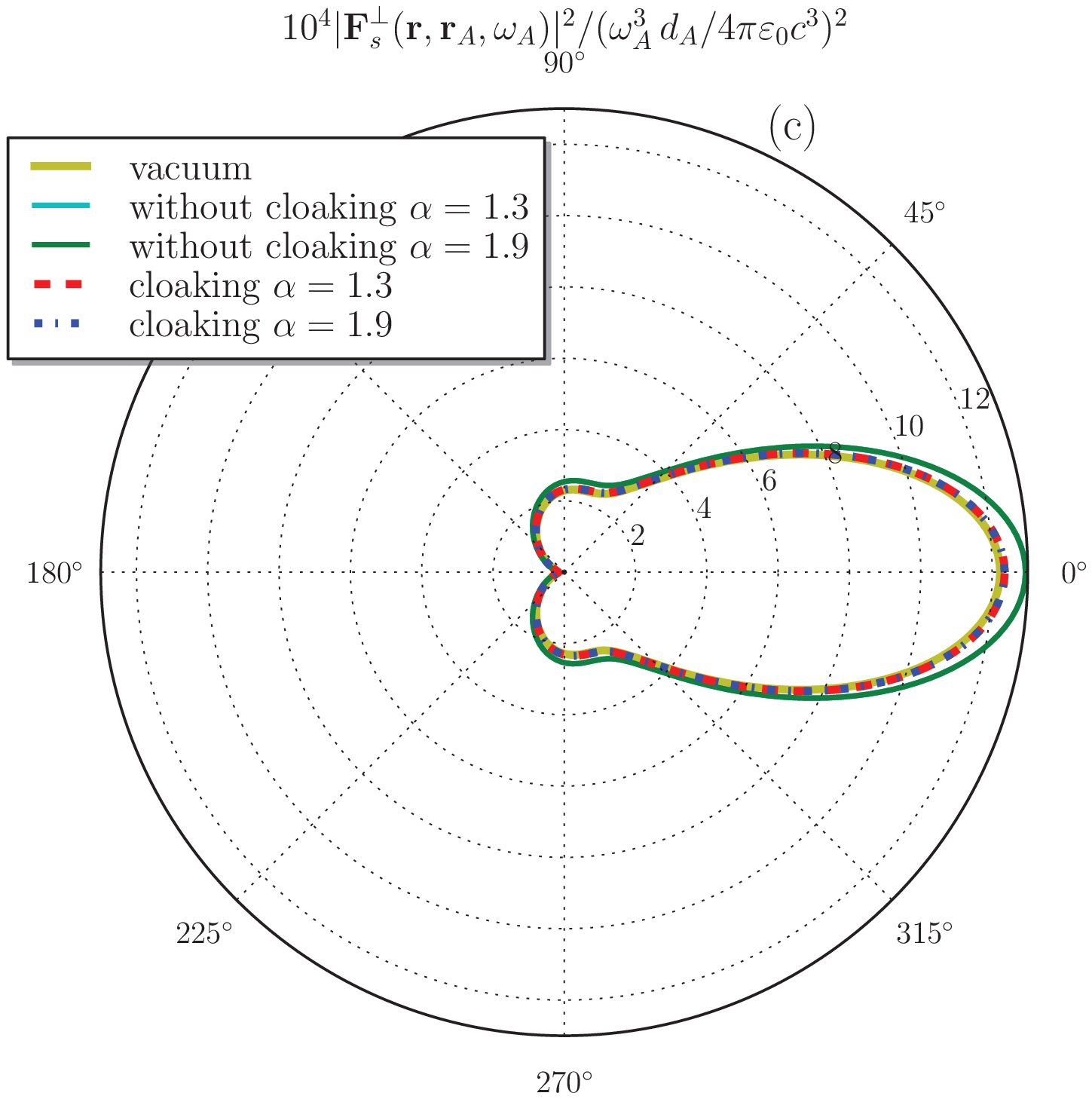}
\end{minipage}
\hspace{0 cm}
\begin{minipage}[b]{0.49\linewidth}
\centering
\includegraphics[width=7.5cm,height=7.5cm]{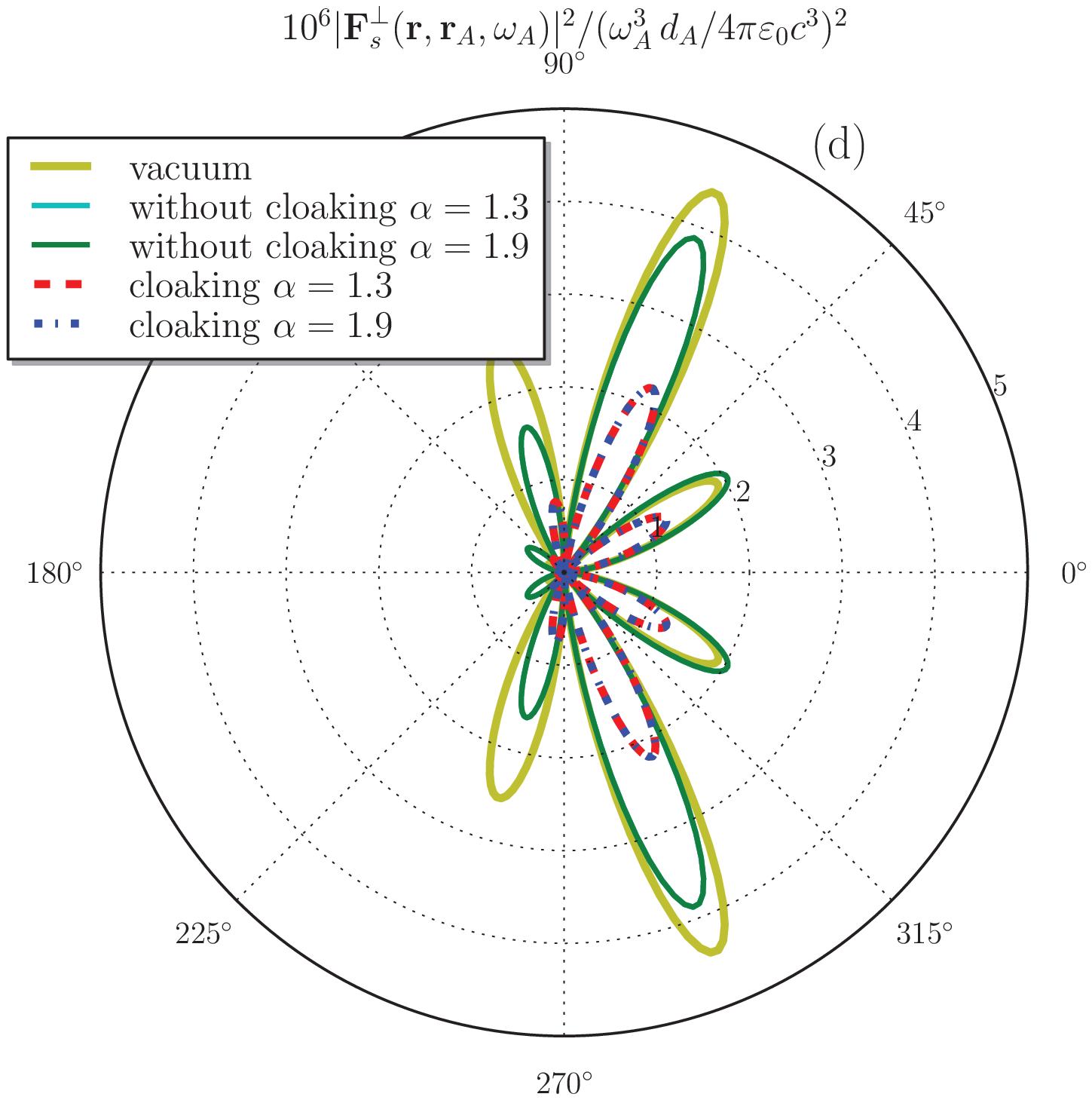}
\end{minipage}
\caption{ Polar diagram of the normalized far-field emission pattern $ \vert {\bf
F}_{w}\left( {{\bf r},{{\bf r}_A},\omega } \right) \vert ^{2}\Big
/(k_{A}^{3} {{d}}_{A} / 4\pi \varepsilon_{0})^{2}$ for a radially oriented transition dipole moment that is placed at $ r_{A} = 4.7 c/\omega_{0}$ and the emission intensity is measured at the
location $ r = 20c/\omega_{0} $ apart from the cloak. The left and right panels correspond to nonresonant ($ \omega=0.01\omega_{0}$) and resonant  ($ \omega=\omega_{0} $) interaction of the excited atom with the field, respectively. Upper panels are for weak coupling regime and lower panels for the strong coupling regime. The size of the cloak
and the material parameters of the cloaking device are identical to those used in Fig.~\ref{Fig:layered}. We have chosen $\delta {\omega_c}=0.01\omega_0$. }\label{Fig:spacial}
\end{figure*}
%
%
Note that Eq.~(\ref{intensity2}) is valid for both weak and strong coupling regimes.
In the weak coupling regime, we can apply the Markov approximation and replace $ C_{u}(t')$ by $ C_{u}(t)$. Thus, Eq.(\ref{intensity2}) simplifies to
\begin{eqnarray}\label{intensity weak}
I_w\left( {{\bf r},t} \right) = {\left| {{{\bf F}_w}\left( {{\bf r},{{\bf
r}_A},\omega } \right)} \right|^2}{e^{ - \Gamma t}},
\end{eqnarray}
where the spatial distribution function $ {{\bf F}_w}\left( {{\bf r},{{\bf
r}_A},\omega } \right) $ is defined as
\begin{eqnarray}\label{1weak spacial distribution function}
{{\bf F}_w}\left( {{\bf r},{{\bf r}_A},\omega } \right) &=& -\frac{{ i
\omega _A ^2{{{\mbox{d}}}_A}}}{{{c^{2}\varepsilon _0}}}\bigg[ {\bar{ \bar G}}
\left( {{\bf r},{{\bf r}_A},\omega } \right)\nonumber \\ &-&
\frac{P}{\pi }\int_0^\infty  {\mbox{d}\omega \frac{{\rm Im}{\bar{ \bar G}}(
{{\bf{ r}} ,{{\bf r}_A},\omega })}{{{\omega  + {\omega _A}}}}}\bigg ].
\end{eqnarray}
For the strong atom-field coupling, Eq.~(\ref{intensity2}) on
the basis of Eq.~(\ref{c u strong}) changes to
\begin{eqnarray}\label{intensity strong}
{I_s}\left( {{\bf{r}},t} \right) = {\left| {{{\bf{F}}_s}\left( {{\bf{r}},{{\bf{r}}_A},\omega } \right)} \right|^2}{e^{ - 2\delta {\omega _{c{\mkern 1mu} {\mkern 1mu} }}t}}{\sin ^2}\left( {\frac{{\Omega t}}{2}} \right),
\end{eqnarray}
where the spatial distribution function $ {{\bf F}_s}\left( {{\bf r},{{\bf
r}_A},\omega } \right) $ is given by
\begin{eqnarray}\label{strong spacial distribution function}
{{\bf{F}}_s}\left( {{\bf{r}},{{\bf{r}}_A},\omega } \right) = - \frac{{ i\omega _A^2{\mbox{d}_A}}}{{\pi {c^2}{\varepsilon _0}}}\frac{\delta {\omega _c}}{\Omega} {\rm Im} \,\left[ {\bar{ \bar G}}({{\bf{ r}} ,{{\bf r}_A},\omega_c })\right]. \nonumber\\
\end{eqnarray}
In the above equations, the subscripts $w$ and $s$ refer to the case where the excited atom is weakly and strongly coupled to the
field, respectively.

Let us restrict our attention to a radially oriented transition dipole moment.
In this case, by simply substituting Eq.~(\ref{tangential Green}) into Eq.~(\ref{1weak spacial distribution function}), we arrive at
\begin{subequations}\label{weak spacial distribution function}
\begin{eqnarray}
\hspace{-2cm}{\bf{F}}^{\bot}_{w}\left( {{\bf{r}},{{\bf{r}}_A},\omega } \right) &=& \frac{{k_A^3{{\mbox{d}}_{{A}}}}}{{4\pi {\varepsilon _0}}}\sum\limits_{n = 1}^\infty
{\frac{{(2n + 1)}}{{{k_A}{{{r}}_A}}}}
\nonumber \\
&\times& \left[ {{j_n}\left( {{k_A}{{{r}}_A}} \right) + B_N^{11}h_n^{(1)}\left( {{k_A}{{{r}}_A}} \right)} \right]\nonumber \\
&\times&
\left[ {{{\bf e}_r}\frac{{n(n + 1)h_n^{(1)}\left( {{k_A}{{r}}} \right)}}{{{k_A}r}}{P_n}\left( {\cos \theta } \right)} \right.{\mkern 1mu} \nonumber
\\
&-&\left. {{{\bf{e}}_\theta }\frac{{\left[ {{k_A}rh_n^{(1)}{{k_A}r}} \right]^\prime}}{{\left( {{k_A}r} \right)}}\sin\theta {P}^\prime _n (\cos\theta)} \right],\,\,\,\,\,\,\,\, \\
{\bf{F}}^{\bot}_{s}\left( {{\bf{r}},{{\bf{r}}_A},\omega } \right) &=&{\delta {\omega _c}}/{\Omega}\,\, {\rm Im}[{\bf{F}}^{\bot}_{w}\left( {{\bf{r}},{{\bf{r}}_A},\omega } \right)].
\end{eqnarray}
\end{subequations}
%
%

The far-field contribution of the emission pattern $ \vert {\bf
F}^\bot\left( {{\bf r},{{\bf r}_A},\omega } \right) \vert ^{2}\Big
/(k_{A}^{3} {\mbox{d}}_{A} / 4\pi \varepsilon_{0})^{2}$ for a radially oriented transition dipole moment in weak and strong coupling regimes is plotted in Fig.~\ref{Fig:spacial} for two cases: at the cloak resonance $ {\omega}={\omega_{0}} $ and far from the cloak resonance ${\omega}=0.01{\omega_{0}}$ The cyan and the green curves represent the influence of the central hidden object with different material parameters $\alpha=1.3$ and $\alpha=1.9$ on the intensity of the spontaneously emitted light, while the results for the combination of the cloak and object are depicted by the blue dashed and red dash-dotted curves. For more clarification, our results are compared with the yellow curve associated with the spontaneously emission pattern of the excited atom in free space.
%

%

In Fig.~\ref{Fig:spacial}, the emission patterns exhibit a symmetry with respect to the plane $\theta=0$ because of our symmetry system.
In part (a) of this figure, far from the cloak resonance frequency and in the weak coupling regime, they give rise to an approximately one-lobe structure which the most pronounced amounts of the intensity placed in the interval $ 0<\theta<45^\circ $. We observe that the emission patterns in presence of the cloaking shell and independent of the object coincide with the corresponding plot in free apace, while in absence of the cloak the emission patterns associated to the object with material parameter $\alpha= 1.3 $ and $\alpha= 1.9 $ are not only mismatch with each other but also with the plot in free apace.

%

Within the cloak resonance frequency, a strikingly different behavior is observed and the pattern of emission changes to a two-lobe structure, as it is seen in Fig.~\ref{Fig:spacial}(b). Obviously, a photon that is spontaneously emitted is almost
certainly absorbed in resonant interaction and does not contribute to the far field. Consequently, the emission intensity decreases. However, the absorbtion is substantial in presence of the combination of the cloak and the object such that the emission intensity independent of the objects reduces about in the order of $2$, because of the emitted light is in resonance with both the cloak shell and the object.
Furthermore, we observe that the emission patterns bent away from the cloak shell as compared to that case in free space. Although, this inclination behavior is lesser when the cloak is absent.

%

In Fig.~\ref{Fig:spacial}(c),
the emission patterns in the strong coupling regime and far from the resonance frequency resembles that observed for nonresonant interaction in the weak coupling regime (Fig.~\ref{Fig:spacial}(a)), but the emission intensity is reduced by magnitude order of $10^3$, since, the absorption losses have a more prominent role in strong coupling regime. It is seen that the cyan and green curves in strong couplin regime, which is associated to different hidden object $\alpha=1.3$ and $\alpha=1.9$, coincide with each other such that a slight difference remains between plots with and without the cloak near $\theta=0$. Therefore, the similarity between the plots far from the cloak resonance frequency is fairly good even in the strong regime and the previous results obtained by studying the spontaneous emission rate are achieved.

%
The interpretation of the curves in Fig.~\ref{Fig:spacial}(d) is quite
similar to that of the plots in Fig.~\ref{Fig:spacial}(b). Of course, some differences are also observed: Two-lobe structure changes into a several lobe structure and unlike to that observed in free space some unwanted radiation is appeared near $\theta=\pi$ when the cloak shell is absent. It seams that the performance of the combination of the cloak and the object is fairly good in this direction. Thus, the effect of the cloaking device in the resonance cloak frequency and strong coupling regime reveals that the object becomes more visible.


\section{Conclusion}

We have developed a formalism for studying the quantum features of an invisibility metamaterials cloak. We have given a systematic analysis of spontaneous decay of an excited two-level atom placed in the vicinity of the spherical cloak. For this purpose, we have extended the canonical quantization scheme which was presented in~\cite{Kheirandish 2010}-~\cite{Amooghorban 2016}, to the case that the electromagnetic field interact with charge particles in the presence of the dispersive, absorptive, anisotropic and inhomogeneous magnetodielectric medium.

%

Then we have used this rigorous formalism to investigate the spontaneous decay of the excited atom near a spherical cloak surrounded the hidden objects in two regimes: weak coupling in which the excited atomic state decays exponentially, and the strong coupling regime which is realized when the atomic transition frequency is close to the cloak resonance frequency. Since, the spontaneous emission is stated in terms of the electromagnetic Green tensor, we have extracted the Green tensor via two methods: exact and discrete methods. The former is directly computed for our system, while the latter one calculated by modeling the cloak shell through concentric layered structure of thin and extending the relations in~\cite{Tai 1994}-~\cite{Qiu 2007}.

In study of the spontaneous decay rate of the excited atom near the cloaking device, we have assumed that the material absorbtion and dispersion of the cloak and also the central object is modeled by Lorentz model. It is seen that the spontaneous decay rate is in agreement with that case in free space  at moderate distance and far from the cloak resonance frequency and thus the cloak shell works well to conceal the object. Moreover, we observed that the performance of the cloak is independent of the central object.

Further investigations are also necessary in order to give a detailed analysis of metamaterial invisibility cloak in the quantum mechanic domain. So, we have discussed the spatial distribution of the spontaneously emitted light in the weak and strong coupling regime. Far from the cloak resonance frequency and within both weak and strong coupling regimes, it is seen that the emission patterns in the presence of the cloak coincide with the corresponding plot in free space. Therefore, the performance of the cloak shell to render the object invisible is good even in strong regime. At near the resonance frequency, the cloak is highly dissipative and dispersive so that the emission pattern takes a form that is almost different from the vacuum pattern which causes the hidden object became more 
visible than the results observed in the classical domain.\\

\section*{Acknowledgment}
M. Morshed Behbahani, E. Amooghorban and A. Mahdifar wish to thank the Shahrekord University for their support.

\appendix
%
\section{Green tensor of spherical invisibility cloaking}\label{Appen:Green tensor}
In this appendix, we extract the electromagnetic Green tensor for an inhomogeneous, anisotropic and dispersive spherical cloak shell which enclosed a homogeneous and isotropic medium as a central object should be hidden.

Considering Eqs.~(\ref{Heisenberg eq of A & E}) and the constitutive relations followed after it, the Maxwell equations for a source-free medium are given by
\begin{subequations}\label{Maxwell}
\begin{eqnarray}
{\bf{\nabla}}\times [(\varepsilon_{0}{{\bar{\bar \varepsilon}}})^{-1}\cdot{\bf {D}}]
= -i\omega{\bf {B}},\label{Maxwell1}\\
{\bf{\nabla}}\times [(\mu_{0}{{\bar{\bar \mu}}})^{-1}\cdot{\bf {B}}]= i\omega{\bf {D}}\label{Maxwell2}.
\end{eqnarray}
\end{subequations}
We decompose the fields into TE and TM modes with respect to $\hat{r}$ by introducing
the scalar potentials as
\begin{subequations}\label{scalar potential}
\begin{eqnarray}
{\bf{B}}_{TM}&=& {\bf{\nabla}}\times(\hat{r}\psi_{TM}),\label{B+scalar potential}\\
{\bf{D}}_{TE}&=& -{\bf{\nabla}}\times(\hat{r}\psi_{TE}),\label{D+scalar potential}\\
{\bf B}_{TE}&=& \frac{1}{i\omega}\bigg\{\nabla\times \left[
{(\varepsilon_{0}{{\bar{\bar \varepsilon}}})^{-1}}\cdot{\nabla}\times (\hat{\bf
r}\psi_{TE})\right] \bigg\},\label{B TE}\\
{\bf D}_{TM}&=& \frac{1}{i\omega}\bigg\{\nabla\times \left[
{(\mu_{0}{{\bar{\bar \mu}}})^{-1}}\cdot{\nabla}\times (\hat{\bf
r}\psi_{TM})\right] \bigg\}.\label{D TM}
\end{eqnarray}
\end{subequations}
By inserting Eqs.~(\ref{D+scalar potential}) and~(\ref{B TE}) into the Maxwell equation~(\ref{Maxwell2}) and equating the radial components,
we can get the wave equation for the scalar potential $ \psi_{TE} $ as follows:
\begin{eqnarray}\label{del1}
\frac{\mu_{r}}{\mu_{t}}\frac{\partial^{2}\psi_{TE}}{\partial
r^{2}}+\nabla^{2}_{t}\psi_{TE}+\omega^{2}\mu_{0}\varepsilon_{0}\mu_{t}\varepsilon_{r}
\psi_{TE}=0.
\end{eqnarray}
where
\begin{eqnarray}\label{laplace operater}
\nabla^{2}_{t}=
\frac{1}{r^{2}\sin\theta}\frac{\partial}{\partial\theta}\Big(\sin\theta\frac{\partial}{\partial\theta}
\Big)+ \frac{1}{r^{2}\sin^{2}\theta}\frac{\partial^{2}}{\partial
\phi^{2}}.
\end{eqnarray}
In a similar way, the wave equation for the scalar potential $ \psi_{TM} $ is obtained by substituting
Eqs.~(\ref{B+scalar potential}) and~(\ref{D TM}) into Eq.~(\ref{Maxwell1}) as
\begin{eqnarray}\label{del2}
\frac{\varepsilon_{r}}{\varepsilon_{t}}\frac{\partial^{2}\psi_{TM}}{\partial
r^{2}}+\nabla^{2}_{t}\psi_{TM}+\omega^{2}\mu_{0}\varepsilon_{0}\mu_{t}\varepsilon_{r}
\psi_{TM}=0.
\end{eqnarray}
An interesting thing arises when the cloak parameters~(\ref{permittivity & permeability _t & r}) are inserted into above equations. In this case, we have: $\frac{\mu_{t}}{\mu_{r}}=\frac{\varepsilon_{t}}{\varepsilon_{r}}= (\frac{r}{r-b})^{2}
$. Therefore, the wave equations~(\ref{del1}) and~(\ref{del2}) for both scalar potentials $ \psi_{TE} $ and $ \psi_{TM} $ become identical. Hereafter, we drop the subscript TE and TM, and specify the scalar potential by $ \psi $.
To solve the scalar wave equation, we use the separation of variable method and assuming $ \psi=
f(r)g(\theta)h(\phi) $.  We find that $ g(\theta) $ and
$ h(\phi) $ are, respectively, associated Legendre polynomials and
harmonic functions, and $ f(r) $ is the solution
of the following equation:
\begin{eqnarray}\label{f(r)}
\left\lbrace   \frac{\partial^{2}}{\partial r^{2}}+\left[
k_{t}^{2}-\beta\frac{n(n+1)}{r^{2}}\right] \right\rbrace f(r)=0,
\end{eqnarray}
where $ k_{t}=\frac{\omega}{c}\sqrt {{\varepsilon _t}{\mu _t}} $ and
$
\beta=\frac{\mu_{t}}{\mu_{r}}=\frac{\varepsilon_{t}}{\varepsilon_{r}}
$ is anisotropic ratio of the cloak.
We will solve the above equation in two ways: \textit{exact} method and
\textit{discrete} method. In the following, we briefly illustrate these methods.
%
%
\subsection{Exact method}
By using the relation between material parameters
specified in~(\ref{permittivity & permeability _t & r}), {\rm i.e. }$ \beta=(\frac{r}{r-b})^{2}$, Eq.~(\ref{f(r)}) in the cloak shell is converted to the Riccati-Bessel equation
\begin{eqnarray}\label{fm}
\left\lbrace   \frac{\partial^{2}}{\partial r^{2}}+\left[
k_{t}^{2}-\frac{n(n+1)}{(r-b)^{2}}\right] \right\rbrace f(r)=0.
\end{eqnarray}
The solution of the above equation is of the form
\begin{eqnarray}\label{solution of f(r)}
f(r)=k_{t}(r-b) z_{n}^{(l)}\big(k_{t}(r-b)\big),
\end{eqnarray}
where the superscripts $l=0$ and $l=1$ refer to the Riccati-Bessel functions of the first
and the third kind, respectively. In other word, the superscripts $(0)$ and $(1)$ denote that the first-type
spherical Bessel function, $j_{n}$, and the third-type
spherical Bessel function or the first-type spherical Hankel
function, $h_{n}^{(1)}$, should be chosen in Eq.~(\ref{solution of f(r)}).
From the above analysis, we find the general solution of Eq.~(\ref{del1}) in the cloak
shell
\begin{eqnarray}\label{Ricatti-Bessel functions}
\psi^{cl}&=&\sum_{m,n}a_{m,n}k_{t}(r-b)
j_n\big({k_t}(r-b)\big) \nonumber \\
&\times& P^{m}_{n}(\cos\theta)\left( {\begin{array}{*{20}{c}}
   {\cos }  \\
   {\sin}  \\
\end{array}} \right)m\phi.
\end{eqnarray}
Let us consider TE/TM decomposition~(\ref{scalar potential}) and the scalar potential~(\ref{Ricatti-Bessel functions}), the electromagnetic fields in the cloak shell can therefore be rewritten as
\begin{subequations}\label{E and H TM&TE}
\begin{eqnarray}
{\bf{E}}^{cl}&=&{\bf{E}}^{cl}_{TE}+{\bf{E}}^{cl}_{TM} \nonumber \\
&=&\sum_{m,n}D_{m,n}\, a_{m,n} \Big[\frac{k_{t}}{\varepsilon_{0}\varepsilon_{t}} M_{{}_o^emn}^{\left( cl \right)}\left( {{k_t}} \right)+\frac{\omega}{i} N_{{}_o^emn}^{\left( cl \right)}\left( {{k_t}} \right)\Big],\label{E TM&TE}\nonumber \\ \\
{\bf{H}}^{cl}&=&{\bf{H}}^{cl}_{TE}+{\bf{H}}^{cl}_{TM} \nonumber \\
&=& -i \sqrt{\frac{\varepsilon_{t}}{\mu_{t}}}\sum_{m,n}D_{m,n}\, a_{m,n} \Big[ M_{{}_o^emn}^{\left( cl \right)}\left( {{k_t}} \right)+ N_{{}_o^emn}^{\left( cl \right)}\left( {{k_t}} \right)\Big]\label{H TM&TE}.\nonumber \\
\end{eqnarray}
\end{subequations}
where $D_{m,n}={\left( {2 - \delta _m^0} \right)} \frac{{2n + 1}}{{n\left( {n + 1} \right)}}\frac{{\left( {n - m} \right)!}}{{\left( {n + m} \right)!}}\label{D_{m,n}}$, $ M_{{}_o^emn}^{\left( cl \right)} $ and $ N_{{}_o^emn}^{\left( cl \right)} $ are the modified spherical vector wave functions corresponding to the cloak region, $b<r<a$, and defined as
\begin{subequations}\label{Mcl&Ncl}
\begin{eqnarray}
M_{{}_o^emn}^{\left( cl \right)}\left( {{k_t}} \right) &=& \frac{r-b}{r}\bigg[ \mp\frac{m}{\sin\theta}z_{n}^{(l)}({k_t}(r-b))\nonumber \\
&\times &P_n^m\left( {\cos \theta } \right)\left( {\begin{array}{*{20}{c}}
{\sin m\phi }  \\
{\cos m\phi }  \\
\end{array}} \right) \hat{\theta} \nonumber \\
& -& \frac{{\mbox{d}P_n^m\left( {\cos \theta } \right)}}{{\mbox{d}\theta }}z_{n}^{(l)}({k_t}(r-b))\left( {\begin{array}{*{20}{c}}
{\cos m\phi }  \\
{\sin m\phi }  \\
\end{array}} \right) \hat{\phi}\bigg],\label{Mcl} \nonumber \\ \\
N_{{}_o^emn}^{\left( cl \right)}\left( {{k_t}} \right)&=& \bigg[\frac{n(n+1)}{{k_t}(r-b)}z_{n}^{(l)}({k_t}(r-b))\nonumber \\
&\times& P_n^m\left( {\cos \theta } \right)\left( {\begin{array}{*{20}{c}}
   {\cos m\phi }  \\
   {\sin m\phi }  \\
\end{array}} \right)\bigg] \hat{r} \nonumber \\
&+&\frac{1}{kr}\frac{\mbox{d}[(r-b)z_{n}^{(l)}({k_t}(r-b))]}{\mbox{d}r} \nonumber \\
&\times & \bigg[\frac{{\mbox{d}P_n^m\left( {\cos \theta } \right)}}{{\mbox{d}\theta }}\left( {\begin{array}{*{20}{c}}
   {\cos m\phi }  \\
   {\sin m\phi }  \\
\end{array}} \right) \hat{\theta} \nonumber \\
 &\mp & \frac{m}{\sin\theta}P_n^m\left( {\cos \theta } \right)\left( {\begin{array}{*{20}{c}}
   {\sin m\phi }  \\
   {\cos m\phi }  \\
\end{array}} \right) \hat{\varphi} \bigg].\label{Ncl}
\end{eqnarray}
\end{subequations}
%
%
%
Now with the solution~(\ref{Ricatti-Bessel functions}) for the
cloak shell, we can also get the field expressions in the regions $ r>a $ and $ r<b $ [see Fig.~\ref{Fig:model}(a)]. Because these two regions are, respectively, free space and the isotropic and homogenous magnetodielectric medium, the medium parameters consequently does not vary with radius. So, similar to consideration in~\cite{Qiu 2007}, the corresponding spherical vector wave functions can be obtained by simple replacement
\begin{eqnarray}\label{tabdil}
N_{{}_o^emn}^{\left( cl \right)}\left( {{k_t}} \right),M_{{}_o^emn}^{\left( cl \right)}\left( {{k_t}} \right)&\longrightarrow & N_{{}_o^emn}\left( {{k_{1,3}}} \right),M_{{}_o^emn}\left( {{k_{1,3}}} \right), \nonumber\\
k_{t}(r-b)&\longrightarrow & k_{1,3}\,r.
\end{eqnarray}
Here the index $1$ and $3$ refer to the regions $ r>a $ and $ r<b $, respectively. Moreover, we have $ k_{1}= \frac{\omega}{c} $ and $ k_{3}= \frac{\omega}{c}\sqrt{\varepsilon_{3}\mu_{3}} $. With these in mind, the spherical vector wave functions for this two regions are written as
\begin{subequations}\label{M&N 1.3}
\begin{eqnarray}
M_{{}_o^emn}\left( {{k_{1,3}}} \right)&=&\bigg[ \frac{\mp m}{\sin\theta}z_{n}^{(l)}({k_{1,3}}\,r)P_n^m\left( {\cos \theta } \right)\left( {\begin{array}{*{20}{c}}
   {\sin m\phi }  \\
   {\cos m\phi }  \\
\end{array}} \right) \hat{\theta}\nonumber\\
&-&\frac{{\mbox{d}P_n^m\left( {\cos \theta } \right)}}{{\mbox{d}\theta }}z_{n}^{(l)}({k_{1,3}}\,r)\left( {\begin{array}{*{20}{c}}
   {\cos m\phi }  \\
   {\sin m\phi }  \\
\end{array}} \right) \hat{\phi}\bigg], \nonumber \\
\\
N_{{}_o^emn}\left( {{k_{1,3}}} \right)&=&\frac{n(n+1)}{{k_{1,3}}\,r}z_{n}^{(l)}({k_{1,3}}\,r)P_n^m\left( {\cos \theta } \right)\left( {\begin{array}{*{20}{c}}
   {\cos m\phi }  \\
   {\sin m\phi }  \\
\end{array}} \right) \hat{r} \nonumber \\
&+&\frac{1}{kr}\frac{\mbox{d}[r z_{n}^{(l)}({k_{1,3}}\,r)]}{\mbox{d}r} \nonumber \\
&\times& \bigg[\frac{{\mbox{d}P_n^m\left( {\cos \theta } \right)}}{{\mbox{d}\theta }}\left( {\begin{array}{*{20}{c}}
   {\cos m\phi }  \\
   {\sin m\phi }  \\
\end{array}} \right) \hat{\theta}  \hspace{0.5 cm}\nonumber \\
&\mp&\frac{m}{\sin\theta}P_n^m\left( {\cos \theta } \right)\left( {\begin{array}{*{20}{c}}
   {\sin m\phi }  \\
   {\cos m\phi }  \\
\end{array}} \right) \hat{\varphi} \bigg] .
\end{eqnarray}
\end{subequations}
In terms of the spherical vector wave functions specified in~(\ref{Mcl&Ncl}) and~(\ref{M&N 1.3}), the Green tensor of system can be constructed. Making use of the method of scattering superposition~\cite{Tai 1994}, the Green tensor may be separated into two parts as
\begin{eqnarray}\label{Gv&Gs}
{\bar{\bar G}} \left( {{\bf{r}},{\bf{r'}},\omega } \right) =
{\bar{\bar G}} {\,_V}\left( {{\bf{r}},{\bf{r'}},\omega } \right) +{
\bar{\bar G}} {\,_S}\left( {{\bf{r}},{\bf{r'}},\omega } \right),\
\end{eqnarray}
where  $ \bar{\bar G} {\,_V}\left( {\bf r,\bf r',\omega } \right) $ and $ \bar{\bar G} {\,_S}\left( {\bf r,\bf r',\omega } \right) $
represent the unbounded and the scattering Green tensor, respectively. The former describes the contribution due to the source in the
infinite homogeneous space while the latter corresponds the contribution
of the source due to the presence of the cloak interfaces. In this paper, the radiation source is located in the free space (region $1$), therefore, we only require the unbounded Green tensor in the vacuum i.e., $ \bar{\bar G} {\,_0}\left( {\bf r,\bf r',\omega } \right) $. Following the procedure given by Tai~\cite{Tai 1994} for the magnetic and the electric type of dyadic Green function, the unbounded Green tensor under the spherical coordinate system can be expressed for $r\lessgtr r'$ as follows:
\begin{eqnarray}\label{Gv}
{\bar{\bar G}} {\,_0}\left( {{\bf{r}},{\bf{r'}},\omega } \right)
=\frac{- \hat{r}\hat{r}}{{\omega}^{2}\varepsilon_{0}\varepsilon_{r}}\delta(r-r^{\prime})+\frac{ik_{1}\mu_{0}\mu_{1}}{4\pi} \sum_{n=0}^{\infty}\sum_{m=0}^{\infty}
D_{m,n}\nonumber \\
 \hspace{-0.5cm}\times  \left\lbrace  \begin{array}{rl} {M_{_o^emn}^{(1)}({k_1}){\mkern 1mu} M_{_o^emn}^{\prime (1)}({k_1})} + {N_{_o^emn}^{(1)}({k_1}){\mkern 1mu} N_{_o^emn}^{\prime (1)}({k_1})}, \\
r \geq r^{\prime},\nonumber \\
{M_{_o^emn}({k_1}){\mkern 1mu} M_{_o^emn}^{\prime (1)}({k_1})} + {N_{_o^emn}({k_1}){\mkern 1mu} N_{_o^emn}^{\prime (1)}({k_1})},\\
 r \leq r^{\prime}, \nonumber \\
\end{array} \right. \\
\end{eqnarray}
where the prime denotes the coordinates $(r',\theta',\varphi')$ of the
source. With regards to the fact that thet radiation source is embedded inside the free space and taking into account
the multiple transmission and reflection effects, we can construct the scattering Green tensor for $f=1,2,3$ as follows:
\begin{widetext}
\begin{subequations}\label{Green11 and 21 and 31}
\begin{eqnarray}
{\bar{ \bar G}}_s^{(11)}\left( { \bf r,{{\bf r}^\prime },\omega } \right) &=& \frac{{i{k_1}\mu_{0}{\mu_1}}}{{4\pi }}\sum\limits_{n = 0}^\infty  {\sum\limits_{m =
0}^n {{D_{m,n}}} }  \left\{ {\left[ {M_{_o^emn}^{(1)}({k_1})B_M^{11}{\mkern 1mu} M_{_o^emn}^{\prime (1)}({k_1})} \right]}+ \right.  \left. {\left[ {N_{_o^emn}^{(1)}({k_1})B_N^{11}{\mkern 1mu} N_{_o^emn}^{\prime (1)}({k_1})} \right]} \right\},  \label{Green11} \\
{\bar{ \bar G}}_s^{(21)}\left( { \bf r,{{\bf r}^\prime },\omega } \right) &=& \frac{{i{k_1}\mu_{0}{\mu_1}}}{{4\pi }}\sum\limits_{n = 0}^\infty  {\sum\limits_{m =
0}^n {{D_{m,n}}} }  \left\{ {\left[ {M_{_o^emn}^{(cl)}({k_t})B_M^{21}{\mkern 1mu} M_{_o^emn}^{\prime (1)}({k_1})} \right]}+ \right.  \left. {\left[ {N_{_o^emn}^{(cl)}({k_t})B_N^{21}{\mkern 1mu} N_{_o^emn}^{\prime (1)}({k_1})} \right]} \right\}\label{Green21} \nonumber \\
&+& \left[ M^{(cl)}_{{}_o^emn} ({k_t}) D_M^{21}\, M^{\prime (1)}_{{}_o^emn}({k_1}) \right] +\left[ N^{(cl)}_{{}_o^emn} ({k_t}) D_N^{21}\, N^{\prime (1)}_{{}_o^emn}({k_1}) \right], \\
{\bar{ \bar G}}_s^{(31)}\left( { \bf r,{{\bf r}^\prime },\omega } \right) &=& \frac{{i{k_1}\mu_{0}{\mu_1}}}{{4\pi }}\sum\limits_{n = 0}^\infty  {\sum\limits_{m =
0}^n {{D_{m,n}}} }  \left\{ {\left[ {M_{_o^emn}({k_3})D_M^{31}{\mkern 1mu} M_{_o^emn}^{\prime (1)}({k_1})} \right]}+ \right.  \left. {\left[ {N_{_o^emn}({k_3})D_N^{31}{\mkern 1mu} N_{_o^emn}^{\prime (1)}({k_1})} \right]} \right\},\label{Green31}
\end{eqnarray}
\end{subequations}
%
where the $ B_{M,N} $ and $ D_{M,N} $ are the coefficients of the scattered Green tensor to be determined. To determined the unknown coefficients in~(\ref{Green11 and 21 and 31}), the boundary conditions satisfied by the Green tensor in the spherical interfaces should be used. These boundary conditions at the spherical interfaces $ r=a $ and $ r=b $ are given as following to ensure continuity of tangential electric and magnetic fields:
\begin{subequations}\label{Boundary conditions}
\begin{eqnarray}
\hat{r}\times \left[ {\bar{\bar G}} {\,_0} +{\bar{\bar G}}_{s}^{(11)}\right]\Bigg|_ {r = b}&=&\hat{r}\times {\bar{\bar G}}_{s}^{(21)}\,\,\,\,\,\Bigg|_{r = b},\\
\hat{r}\times {\bar{\bar G}}_{s}^{(21)}\Bigg|_ {r =a}&=&\hat{r}\times {\bar{\bar G}}_{s}^{(31)} \Bigg|_{r =a},
\\
\frac{1}{\mu_{1}}\hat{r}\times {\nabla}\times \left[  {\bar{\bar G}} {\,_0} +{\bar{\bar G}}_{s}^{(11)}\right] \Bigg|_{r =b} &=&\hat{r}\times \left[ {\bar{\bar{\mu}}_{2}^{-1}}\cdot {\nabla}\times{\bar{\bar G}}_{s}^{(21)}\right] \Bigg|_ {r =b}, \qquad  \\
\hat{r}\times \left[ {\bar{\bar{\mu}}_{2}^{-1}}\cdot {\nabla}\times{\bar{\bar G}}_{s}^{(21)}\right] \Bigg|_ {r =a}&=&\frac{1}{\mu_{3}}\hat{r}\times {\nabla}\times{\bar{\bar G}}_{s}^{(31)} \Bigg|_{r =a}.
\end{eqnarray}
\end{subequations}
\end{widetext}
In this paper only the scattering coefficients $ B_{M,N}^{11}$ are of interest and must be determined. Since the decay rate of the excited atom, which is located in free space in our case, depends on the Green tensor of system at the position of the atom, therefore, the source and the field points are located in region $1$. By inserting Eq.~(\ref{Green11 and 21 and 31}) into Eq.~(\ref{Boundary conditions}) and solving the obtained equations, the unknown coefficients $ B_{M,N}^{11}$ are derived as
\begin{eqnarray}\label{B_MN}
B_{M,N}^{11}=- \frac{{T_{P1}^{H,V}R_{F1}^{H,V} +
T_{F1}^{H,V}R_{F2}^{H,V}}}{{ R_{F2}^{H,V} R_{P1}^{H,V} T_{F1}^{H,V}
+ T_{P1}^{H,V}}},
\end{eqnarray}
where the superscript $F$ and $P$ stand for the centrifugal and centripetal
waves and the TE and TM waves represented by the subscripts $H$ and $V$, respectively. The reflection and transition coefficients $R^{H,V}_{F,P}$ and $T^{H,V}_{F,P}$ introduced in Eq.~(\ref{B_MN}) are defined as
\begin{widetext}
\begin{eqnarray}\label{R&T}
T_{F1}^H &=& \frac{{{\eta _1}{\mu _1}{k_2}\left( {\partial {\Im
_2}{\hbar _2} - {\Im _2}\partial {\hbar _2}} \right)}}{{{\mu
_1}{k_2}\partial {\Im _2}{\hbar _1} - {\mu _2}{k_1}{\Im _2}\partial
{\hbar _1}}}, \hspace{1.5cm}
 T_{F1}^V = \frac{{{\eta _1}{\mu
_1}{k_2}\left( {{\Im _2}\partial {\hbar _2} - \partial {\Im
_2}{\hbar _2}} \right)}}{{{\mu _1}{k_2}{\Im _2}\partial {\hbar _1} -
{\mu _2}{k_1}\partial {\Im _2}{\hbar _1}}},\nonumber \\
T_{P1}^V
&=& \frac{{{\eta _1}{\mu _1}{k_2}\left( {\partial {\Im _2}{\hbar _2}
- {\Im _2}\partial {\hbar _2}} \right)}}{{{\mu _1}{k_2}\partial {\Im
_1}{\hbar _2} - {\mu _2}{k_1}{\Im _1}\partial {\hbar _2}}}, \hspace{1.5cm}
T_{P1}^H = \frac{{{\eta _1}{\mu _1}{k_2}\left( {\partial {\Im _2}
{\hbar _2} - \partial {\Im _2}{\hbar _2}} \right)}}{{{\mu
_1}{k_2}{\Im _1}\partial {\hbar _2} - {\mu _2}{k_1}\partial {\Im
_1}{\hbar _2}}},\nonumber \\
R_{F2}^H &=& \frac{{{\mu
_2}{k_3}\partial {\Im _3}{\Im _2} - {\mu _3}{k_2}\partial {\Im
_2}{\Im _3}}}{{{\mu _2}{k_3}\partial {\Im _3}{\hbar _2} - {\mu
_3}{k_2}{\Im _3}\partial {\hbar _2}}}, \hspace{1.3cm}
 R_{F2}^V = \frac{{{\mu
_2}{k_3}{\Im _3}\partial {\Im _2} - {\mu _3}{k_2}{\Im _2}\partial
{\Im _3}}}{{{\mu _2}{k_3}{\Im _3}\partial {\hbar _2} - {\mu
_3}{k_2}\partial {\Im _3}{\hbar _2}}},\\
R_{F1}^V &=&
\frac{{{\mu _1}{k_2}{\Im _2}\partial {\Im _1} - {\mu _2}{k_1}{\Im
_1}\partial {\Im _2}}}{{{\mu _1}{k_2}{\Im _2}\partial {\hbar _1} -
{\mu _2}{k_1}\partial {\Im _2}{\hbar _1}}}, \hspace{1.3cm}
R_{F1}^H =
\frac{{{\mu _1}{k_2}\partial {\Im _2}{\Im _1} - {\mu
_2}{k_1}\partial {\Im _2}{\Im _1}}}{{{\mu _1}{k_2}\partial {\Im
_2}{\hbar _1} - {\mu _2}{k_1}{\Im _2}\partial {\hbar _1}}},\nonumber \\
 R_{P1}^H &=& \frac{{{\mu _1}{k_2}\partial {\hbar _2}{\hbar
_1} - {\mu _2}{k_1}{\hbar _2}\partial {\hbar _1}}}{{{\mu
_1}{k_2}\partial {\hbar _2}{\Im _1} - {\mu _2}{k_1}{\hbar
_2}\partial {\Im _1}}},\hspace{1.49cm}
R_{P1}^V = \frac{{{\mu _1}{k_2}\partial
{\hbar _1}{\hbar _2} - {\mu _2}{k_1}{\hbar _1}\partial {\hbar
_2}}}{{{\mu _1}{k_2}{\hbar _2}\partial {\Im _1} - {\mu
_2}{k_1}\partial {\hbar _2}{\Im _1}}}.\nonumber
\end{eqnarray}
In Eq.~(\ref{R&T}), the following parameters have been used to simplify the symbolic calculations
\begin{eqnarray}\label{hankel cl}
{\Im _{1}} &=& {j_{n}}\left( {{k_1}{r}} \right),\hspace{2.65cm}
{\Im _{t}} = {j_{n}}( {k_t}{({r-b)}}),\nonumber \\  {\hbar _{1}} &=&
h_n^{\left( 1 \right)}\left( {{k_t}{r}}
\right),\hspace{2.53cm}
{\hbar
_{t}} = h_n^{\left( 1 \right)}\left( {k_t}{({r-b)}} \right),\nonumber\\
\partial {\Im_{1}} &=& \frac{1}{\rho }\frac{{\mbox{d}\left[ {\rho {j_n}\left( \rho  \right)} \right]}}{{\mbox{d}\rho }}{|_{\rho  = {k_1}{r}}},
\hspace{1.00cm}
\partial {\Im _{t}} = \frac{1}{\rho }\frac{{\mbox{d}\left[ {\rho {j_n}\left( \rho  \right)} \right]}}{{\mbox{d}\rho }}{|_{\rho  = {k_t}({r-b)}}},\\
 \qquad\partial{\hbar _{1}} &=& \frac{1}{\rho
}\frac{{\mbox{d}\left[ {\rho h_n^{\left( 1 \right)}\left( \rho  \right)}
\right]}}{{\mbox{d}\rho }}{|_{\rho  = {k_1}{r}}},\nonumber
\qquad\partial{\hbar_{t}} = \frac{1}{\rho }\frac{{\mbox{d}\left[ {\rho h_n^{\left( 1 \right)}\left( \rho  \right)} \right]}}{{\mbox{d}\rho }}{|_{\rho  =
{k_t}({r-b)}}}.
\end{eqnarray}
\end{widetext}
%
%
\subsection{Discrete method }
%
\begin{table}\label{Table}
\caption{Radial components of permittivity and permeability of each layer for $22$ and $14$ multilayered cloak. The material absorbtion
and dispersion of each layer is described by the Lorentz model with parameters that are identical to those used in Fig.~\ref{Fig:layered}.} \label{Table}
\centering
\begin{tabular}{|c || c | c|} 
\hline

&\multicolumn{2}{|c|}{ $(\mu_r = \varepsilon_r)/{\kappa _L}$}  \\ \hline

$f$ &$N=14$ &$N=22$\\ \hline

1 &1.000&1.000 \\ \hline

2 & 0.330& 0.330  \\ \hline

3& 0.300 & 0.310   \\ \hline

4& 0.260 & 0.290  \\ \hline

5& 0.220 & 0.270  \\ \hline

6& 0.190 & 0.240  \\ \hline

7& 0.150 & 0.220  \\ \hline

8& 0.120 & 0.200  \\ \hline

9& 0.090 & 0.180  \\ \hline

10& 0.060 & 0.160  \\ \hline

11& 0.040 & 0.140  \\ \hline

12& 0.180 & 0.120  \\ \hline

13& 0.005 & 0.100  \\ \hline

14& 1.300 & 0.080  \\ \hline

15 &  $-$ & 0.070 \\ \hline

16&  $-$ & 0.050  \\ \hline

17& $-$ & 0.040  \\ \hline

18 & $-$ & 0.020 \\ \hline

19& $-$ & 0.015  \\ \hline

20& $-$ & 0.007  \\ \hline

21& $-$ & 0.001  \\ \hline

22& $-$ & 1.300  \\ \hline
\end{tabular}
\end{table}
As shown in Fig.~\ref{Fig:model}(b), we model the cloak shell by a large number of the spherical thin layers with
constant material parameters whose radial component in each layer increases outward according to Eq.~(\ref{permittivity & permeability _t & r}).
In this case, which is an approximation of the exact cloak, we can choose the material parameters as
\begin{eqnarray}\label{parameters of layered}
&&\hspace{-0.9cm}{\varepsilon _t}\left( \omega  \right) = {\mu _t}\left( \omega
\right) = \frac{{{a}}}{{{a} - {b}}}\kappa_L(\omega),\\
&&\hspace{-0.9cm}{\varepsilon _r}\left( {\omega ,f} \right) = {\mu
_r}\left( {\omega ,f} \right) = \frac{{{a}}}{{{a} -
{b}}}{\left( {\frac{{{r_{f - 1}} - {b}}}{{{r_{f -
1}}}}} \right)^2}\kappa_L(\omega).
\end{eqnarray}
where $r_f$ corresponds to the location of the $f$th layer.
In Table.~\eqref{Table}, the radial component of the anisotropic permittivity and permeability ~(\ref{permittivity & permeability_r}) is shown for two cases which is used in Fig.~\ref{Fig:layered} with $N=14$ and $N=22$ layered cloak.
Now, we insert the material parameters~(\ref{parameters of layered}) in Eq.~(\ref{f(r)}) and equating the expression $ n (n+1)\frac{\varepsilon_{t}}{\mu_{t}} $ by $ \nu(\nu+1) $. We find that the solution of the Eqs.~(\ref{f(r)}) has a similar form according to Eq.~(\ref{solution of layered}), but with a difference that the spherical Bessel function of the $n$th order and as well as the radius function $(r-b)$ replaced by $\nu=\left[n(n+1)\frac{\varepsilon_{t}}{\varepsilon_{r}}+\frac{1}{4}\right]^{1/2}-\frac{1}{2}$ and $r$, respectively.
%
This allow us to write the general solution of Eq.~(\ref{del1}) inside the layered cloak as
\begin{eqnarray}\label{solution of layered}
\hspace{-0.9cm} \psi^{cl}=\sum_{m,n}b_{m,n}\,k_{t}\, r\, z_{\nu}(k_{t}r) P^{m}_{n}(\cos\theta)\left(
{\begin{array}{*{20}{c}}
   {\cos }  \\
   {\sin}  \\
\end{array}} \right)m\phi.
\end{eqnarray}
The analysis similar to that of the electric and magnetic field in Eqs.~(\ref{E and H TM&TE}) shows that the spherical vector wave functions $ M ^{(cl)}_{{}_o^em\nu}\left( {{k_t}} \right) $ and $ N^{(cl)}_{{}_o^em\nu}\left( {{k_t}} \right) $ inside the layered cloak can be specified by Eqs.~(\ref{M&N 1.3}) with the function $z_{n}(k_{t}r)$ replaced by $z_{\nu}(k_{t}r)$. Of course, these functions remain without any changes outside the cloak. Given that the source is located out of the cloak, i.e., $s=1$, the scattering Green tensor for the $f$th layer is expressed as~\cite{Li1994}
\begin{eqnarray}\label{GsT}
&&{\bar{\bar G}} ^{(f1)}_{s}\left( {{\bf{r}},{\bf{r'}},\omega }
\right) = \frac{ik_{t,s}\mu \mu_{t,f}}{4\pi}
\sum_{n=0}^{\infty}\sum_{m=0}^{\infty} D_{m,n}\nonumber \\  &&\times
\Big( (1-\delta^{N}_{f}) {M_{m\nu}^{(1)}}( k_{f}) \left[
(1-\delta^{N}_{f})B^{f1}_{M}{M_{m\nu}^{\prime{(1)}}}( k_{1}) \right]
\nonumber \\  &&+(1-\delta^{N}_{f}){N_{m\nu}^{(1)}}( k_{f})
\left[B^{f1}_{N}{N_{m\nu}^{\prime{(1)}}}( k_{1})
\right] \nonumber \\ &&+(1-\delta^{1}_{f}){M_{m\nu}^{(1)}}( k_{f})
\left[ D^{f1}_{M}{M_{m\nu}^{\prime{(1)}}}( k_{1}) \right] \nonumber
\\&&+(1-\delta^{1}_{f}){N_{mn}^{(1)}}( k_{f}) \left[
D^{f1}_{N}{N_{m\nu}^{\prime{(1)}}}( k_{1}) \right] \Big).
\end{eqnarray}
where the superscripts $f$ denotes the field point located in $f$th region.
Now, all that remains to be done is to determine the unknown coefficients $ B^{f1}_{M,N} $ and $ D^{f1}_{M,N} $. As explained in the direct method, the field point in our model is located in region $1$. Therefore, we only require the unknown coefficient $ B^{11}_{M,N} $. The procedure for deriving the coefficients $ B^{11}_{M,N} $ is similar to that obtained in Eq.~{\ref{B_MN}}. The boundary conditions should be satisfied by the scattering Green tensor at interfaces $ r=r_{f} $ and $ r=r_{f+1} $. Thus, we have
\begin{subequations}\label{Boundary conditions2}
\begin{eqnarray}
\hat r \times \left[ {{\bar{ {\bar G}}{{\mkern 1mu}} _0} + {\bar{ \bar G}}_s^{(fs)}} \right] &= &\hat r \times {\bar {\bar G}}_s^{(f + 1)s} {\mkern 1mu} {\mkern 1mu} {\mkern 1mu}  \\
\hat r \times {\bar{ \bar G}}_s^{(fs)} &=& \hat r \times {\bar {\bar G}}_s^{(f + 1)s}  \\
\mu^{-1}_{f}\hat r \times {\bf \nabla}  \times \left[ {{\bar {\bar G}}{{\mkern 1mu} _0} + {\bar{ \bar G}}_s^{(fs)}} \right]& = &\hat r \times \left[ {{\bar {\bar \mu}} _{f + 1}^{ - 1}\cdot{\bf \nabla}  \times {\bar{ \bar G}}_s^{(f + 1)s}} \right]  \nonumber \\
\\
\hat r \times \left[ {{\bar {\bar \mu}} _f^{ - 1}\cdot{\bf \nabla}  \times {\bar {\bar G}}_s^{(fs)}} \right] &=& \mu^{-1}_{f + 1}\hat r \times {\bf \nabla}  \times {\bar {\bar G}}_s^{(f + 1)s}.\,\,\,\,\,\,\,\,
\end{eqnarray}
\end{subequations}
Let us introduce the following transmission matrix
%
\begin{eqnarray}
\left[ {{T}}_{l,f}\right]=\left[ \begin{array}{cc}
\frac{1}{T_{F,f}^{H,V}} & \frac{R_{F,f}^{H,V}}{T_{F,f}^{H,V}}  \\  \frac{R_{P,f}^{H,V}}{T_{P,f}^{H,V}}   &  \frac{1}{T_{P,f}^{H,V}} \end{array}\right],
\end{eqnarray}
%
where $l=M,N$ and the reflection and transmission coefficients $R_{(F,P)\, f}^{H, V} $ and $T_{(F,P)\, f}^{H, V}$ are given by
\begin{widetext}
\begin{eqnarray}\label{TTT}
T_{F{\kern 1pt} f}^H &=& \frac{{{\mu _{f + 1}}{k_{f + 1}}\left( {\partial {\Im _{f + 1}}{\hbar _{f + 1}} - {\Im _{f + 1}}\partial {\hbar _{f + 1}}} \right)}}{{{\mu _f}{k_{f + 1}}\partial {\Im _{f + 1}}{\hbar _f} - {\mu _{f + 1}}{k_f}{\Im _{f + 1}}\partial {\hbar _f}}},\hspace{1.00cm}
T_{F{\kern 1pt} f}^V = \frac{{{\mu _{f + 1}}{k_{f + 1}}\left( {{\Im _{f + 1}}\partial {\hbar _{f + 1}} - \partial {\Im _{f + 1}}{\hbar _{f + 1}}} \right)}}{{{\mu _f}{k_{f + 1}}{\Im _{f + 1}}\partial {\hbar _f} - {\mu _{f + 1}}{k_f}\partial {\Im _{f + 1}}{\hbar _f}}},\nonumber \\
T_{P{\kern 1pt} f}^V &=& \frac{{{\mu _{f + 1}}{k_{f + 1}}\left( {\partial {\Im _{f + 1}}{\hbar _{f + 1}} - {\Im _{f + 1}}\partial {\hbar _{f + 1}}} \right)}}{{{\mu _f}{k_{f + 1}}\partial {\Im _f}{\hbar _{f + 1}} - {\mu _{f + 1}}{k_f}{\Im _f}\partial {\hbar _{f + 1}}}},\hspace{1.00cm}
T_{P{\kern 1pt} f}^H = \frac{{{\mu _{f + 1}}{k_{f + 1}}\left( {\partial {\Im _{f + 1}}\partial {\hbar _{f + 1}} - \partial {\Im _{f + 1}}{\hbar _{f + 1}}} \right)}}{{{\mu _f}{k_{f + 1}}{\Im _f}\partial {\hbar _{f + 1}} - {\mu _{f + 1}}{k_f}\partial {\Im _f}{\hbar _{f + 1}}}},\nonumber \\
R_{F{\kern 1pt} f}^V &=& \frac{{{\mu _f}{k_{f + 1}}{\Im _{f + 1}}\partial {\Im _f} - {\mu _{f + 1}}{k_f}{\Im _f}\partial {\Im _{f + 1}}}}{{{\mu _1}{k_{f + 1}}{\Im _{f + 1}}\partial {\hbar _f} - {\mu _{f + 1}}{k_f}\partial {\Im _{f + 1}}{\hbar _f}}},\hspace{1.1cm}
R_{Ff}^H = \frac{{{\mu _f}{k_{f + 1}}\partial {\Im _{f + 1}}{\Im _f} - {\mu _{f + 1}}{k_f}\partial {\Im _{f + 1}}{\Im _f}}}{{{\mu _f}{k_{f + 1}}\partial {\Im _{f + 1}}{\hbar _f} - {\mu _{f + 1}}{k_f}{\Im _{f + 1}}\partial {\hbar _f}}},\nonumber \\
R_{P{\kern 1pt} f}^H &=& \frac{{{\mu _f}{k_{f + 1}}\partial {\hbar _{f + 1}}{\hbar _f} - {\mu _{f + 1}}{k_f}{\hbar _{f + 1}}\partial {\hbar _f}}}{{{\mu _f}{k_{f + 1}}\partial {\hbar _{f + 1}}{\Im _f} - {\mu _2}{k_f}{\hbar _2}\partial {\Im _f}}},\hspace{1.3cm}
R_{P{\kern 1pt} f}^V = \frac{{{\mu _f}{k_{f + 1}}\partial {\hbar _f}{\hbar _{f + 1}} - {\mu _{f + 1}}{k_f}{\hbar _f}\partial {\hbar _{f + 1}}}}{{{\mu _f}{k_{f + 1}}{\hbar _{f + 1}}\partial {\Im _f} - {\mu _{f + 1}}{k_f}\partial {\hbar _{f + 1}}{\Im _f}}},\nonumber \\
%
\end{eqnarray}
\end{widetext}
To simplify the complicated representation, the following parameters
\begin{eqnarray}\label{hankel}
{\Im _{f}} &=& {j_{\nu ,f}}\left( {{k_{tf}}{r_f}} \right){|_{\rho  =
{k_{tf}}{r_f}}}, \nonumber\\ {\hbar _{f}} &=& h_n^{\left( 1
\right)}\left( {{k_{tf}}{r_f}} \right){|_{\rho  = {k_{tf}}{r_f}}}\nonumber \\
\partial {\Im _{f}} &=& \frac{1}{\rho }\frac{{\mbox{d}\left[ {\rho {j_\nu}\left( \rho  \right)} \right]}}{{\mbox{d}\rho }}{|_{\rho  = {k_{tf}}{r_f}}},\nonumber \\
\qquad\partial {\hbar _{f}} &=& \frac{1}{\rho }\frac{{\mbox{d}\left[ {\rho h_\nu ^{\left( 1 \right)}\left( \rho  \right)} \right]}}{{\mbox{d}\rho }}{|_{\rho  =
{k_{tf}}{r_f}}},
\end{eqnarray}
are used in Eq.~(\ref{TTT}). From the boundary conditions~(\ref{Boundary conditions2}), we obtain two recurrence matrices relations.
Solving the coupled matrix equations, after some manipulations, the unknown coefficients $B_{M,N}^{11}$ are derived as
\begin{eqnarray}\label{B_MN layer}
B_{l}^{11} =  - \frac{{T_{l,12}^1}}{{T_{l,11}^1}}.
\end{eqnarray}
Here, the following matrix are also used to shorten the above expression
\begin{eqnarray}\label{TK}
\left[ {{T}}_{l}^{K}\right]_{2\times2}&=& \left[ {{T}}_{l,N-1}\right] \left[ {{T}}_{l,N-2}\right]\cdots\left[ {{T}}_{l,K+1}\right]\left[ {{T}}_{l,K}\right] \nonumber \\
&=&\left[ \begin{array}{cc}
{{T}}_{l,11}^{K} & {{T}}_{l,12}^{K} \\   {{T}}_{l,21}^{K}   &  {{T}}_{l,22}^{K} \end{array}\right].
\end{eqnarray}

\end{document}